\documentclass[10pt,a4paper,twoside,twocolumn,english,aps,manuscript,superscriptaddress, aps,manuscript,preprint,showpacs]{revtex4}

\usepackage[LGR,T1]{fontenc}
\usepackage[latin9]{inputenc}
\usepackage{color}
\usepackage{babel}
\usepackage{float}
\usepackage{textcomp}
\usepackage{amsmath}
\usepackage{graphicx}
\usepackage{esint}
\usepackage[unicode=true,pdfusetitle,
 bookmarks=true,bookmarksnumbered=true,bookmarksopen=false,
 breaklinks=true,pdfborder={0 0 0},pdfborderstyle={},backref=false,colorlinks=true]
 {hyperref}
\hypersetup{
 linkcolor=red, citecolor=blue,  urlcolor=blue}

\makeatletter

\DeclareRobustCommand{\greektext}{%
  \fontencoding{LGR}\selectfont\def\encodingdefault{LGR}}
\DeclareRobustCommand{\textgreek}[1]{\leavevmode{\greektext #1}}
\ProvideTextCommand{\~}{LGR}[1]{\char126#1}

\providecommand{\tabularnewline}{\\}

\@ifundefined{textcolor}{}
{%
 \definecolor{BLACK}{gray}{0}
 \definecolor{WHITE}{gray}{1}
 \definecolor{RED}{rgb}{1,0,0}
 \definecolor{GREEN}{rgb}{0,1,0}
 \definecolor{BLUE}{rgb}{0,0,1}
 \definecolor{CYAN}{cmyk}{1,0,0,0}
 \definecolor{MAGENTA}{cmyk}{0,1,0,0}
 \definecolor{YELLOW}{cmyk}{0,0,1,0}
}

\makeatother

\makeatother

\begin{document}
\title{LiZn$_{2}$V$_{3}$O$_{8}$: A new geometrically frustrated cluster
spin-glass }
\author{S. Kundu}
\email{skundu37@gmail.com}

\affiliation{Department of Physics, Indian Institute of Technology Bombay, Powai,
Mumbai 400076, India}
\author{T. Dey}
\affiliation{Experimental Physics VI, Center for Electronic Correlations and Magnetism,
University of Augsburg, 86159 Augsburg, Germany}
\author{A. V. Mahajan}
\affiliation{Department of Physics, Indian Institute of Technology Bombay, Powai,
Mumbai 400076, India}
\author{N. $\mathrm{B\mathrm{\ddot{u}}ttgen}$}
\affiliation{Experimental Physics V, Center for Electronic Correlations and Magnetism,
University of Augsburg, 86159 Augsburg, Germany}
\date{\today}
\begin{abstract}
{\normalsize{}We have investigated the structural and magnetic properties
of a new cubic spinel LiZn$_{2}$V$_{3}$O$_{8}$ (LZVO) through x-ray
diffraction, dc and ac susceptibility, magnetic relaxation, aging,
memory effect, heat capacity and $^{7}$Li nuclear
magnetic resonance (NMR) measurements.
A Curie-Weiss fit of the dc susceptibility $\chi_{\mathrm{dc}}$($\mathit{T}$)
yields a Curie-Weiss temperature $\mathrm{\theta}_{\mathrm{CW}}$
= -185 K. This suggests strong antiferromagnetic (AFM) interactions
among the magnetic vanadium ions. The dc and ac susceptibility data indicate
the spin-glass behavior below a freezing temperature $T_{f}$ $\simeq$
3 K. The frequency dependence of the $T_{f}$ is characterized by
the Vogel-Fulcher law and critical dynamic scaling behavior or power
law. From both fitting, we obtained the value of the characteristic
angular frequency $\omega_{0}$ $\approx$ 3.56$\times$10$^{6}$
Hz, the dynamic exponent }$\mathit{zv}$ $\approx$ 2.65,{\normalsize{}
and the critical time constant $\tau_{0}$ $\approx$ 1.82$\times$10$^{-6}$
s, which falls in the conventional range for typical cluster spin-glass
(CSG) systems. The value of relative shift
in freezing temperature $\delta T_{f}$ $\simeq$ 0.039 supports a
CSG ground states. We also found aging phenomena and memory effects
in LZVO. The asymmetric response of the magnetic relaxation below
$T_{f}$ supports the hierarchical model. Heat capacity data show
no long-range or short-range ordering down to 2 K. Only about 25\%
magnetic entropy change $(\Delta S_{\mathrm{m}})$ signifies the
presence of strong frustration in the system. The $^{7}$Li NMR spectra show a shift and broadening with
decreasing temperature. The spin-lattice and spin-spin relaxation
rates show anomalies due to spin freezing around 3 K as the bulk
magnetization.}{\normalsize\par}
\end{abstract}
\maketitle

\section{Introduction}

\textcolor{black}{Spinel oxides with the general formula AB$_{2}$O$_{4}$
have provided an excellent arena for studying the effects of geometrical
frustration \citep{Greedan2001,Moessner2006} and have seen a surge
of interest in the past decade due to a series of exciting experimental
observations. In the cubic spinel, the B-sites form a corner shared,
3D tetrahedral network like the pyrochlore lattice which is geometrically
frustrated. Geometrical frustration arises when magnetic moments occupying
the B-sites interact antiferromagnetically with each other. With the
intention of unraveling novel magnetic properties arising due to geometric
frustration, we were in the quest for new B-site cubic spinel compounds.
In this category, 3$\mathit{d}$-transition metal oxide LiV$_{2}$O$_{4}$
\citep{Chmaissem1997,Kondo1997,Krimmel2004} (V$^{3+}$: V$^{4+}$=
1:1) is a well-studied system. It shows heavy fermion behavior \citep{Stewart1984}.
But with non-magnetic impurity (Zn/Ti) doping, it shows spin-glass
behavior \citep{Trinkl2000a,Miyoshi2002,Ueda1997}. Recently studied
mixed valent spinel system Zn$_{3}$V$_{3}$O$_{8}$ or {[}Zn$_{1.0}$(Zn$_{0.25}$V$_{0.75}$)$_{2}$O$_{4}${]}$_{2}$
with V$^{3+}$: V$^{4+}$= 2:1 \citep{Chakrabarty2014} has been suggested
to possess a cluster spin-glass ground state. Similarly, Li$_{2}$ZnV$_{3}$O$_{8}$
(better written as {[}Zn$_{0.5}$Li$_{0.5}$(Li$_{0.25}$V$_{0.75}$)$_{2}$O$_{4}${]}$_{2}$)
containing only V$^{4+}$ ($S=1/2$) magnetic ions shows spin-glass behavior at
low temperatures\citep{Chakrabarty2014a}. LiZn$_{2}$Mo$_{3}$O$_{8}$
\citep{Sheckelton2012,Sheckelton2014} is another system stoichiometrically
similar to our probed system. With a frustrated geometry, it exhibits
a resonating valence-bond condensed state \citep{Anderson1973}. }The
strong interplay between the spin, charge, lattice and orbital degrees
of freedom in these transition metal oxides (TMO) are the source of
these novel magnetic properties.\textcolor{black}{{} Motivated by these
exotic behaviors of the cubic spinels, we decided to investigate the
mixed valent system LiZn$_{2}$V$_{3}$O$_{8}$ (LZVO). Only the structural
and electrical properties of LZVO were reported long back in 1972
by B. Reuter and G. Colsmann \citep{Reuter1972}. The magnetic properties
of LZVO have not been reported so far. As zinc prefers the A-site in spinel
like in ZnCr$_{2}$O$_{4}$ \citep{Lee2000,Lee2002} and ZnV$_{2}$O$_{4}$
\citep{Lee2004,Reehuis2003}, we expect the site occupancy of LZVO
to be represented as {[}Zn$_{1.0}$(Li$_{0.25}$V$_{0.75}$)$_{2}$O$_{4}${]}$_{2}$
- where the octahedral B-site is occupied by Li and V in the 1:3 ratio.
It will be interesting to observe how the 25\% dilution via non-magnetic
lithium affects the} magnetic properties of LZVO. Also, this system
is amenable to nuclear magnetic resonance (NMR) local probe measurement of $^{7}$Li nuclei.\textcolor{black}{{}
From stoichiometry, LZVO is a mixed valent spinel as in this compound
the average vanadium valence is $+\frac{11}{3}.$ So it will be interesting
to see whether this leads to tetravalent and trivalent V in a 2:1
ratio.}

The motivation behind studying the system \textcolor{black}{LiZn$_{2}$V$_{3}$O$_{8}$
}is to contrast its properties with those of the resonating valence
bond solid and cluster magnet compound LiZn$_{2}$Mo$_{3}$O$_{8}$,
whereas Zn$_{3}$V$_{3}$O$_{8}$ (ZVO) was motivated by Majumdar-Ghosh
chain system Ba$_{3}$V$_{3}$O$_{8}$. Also, in LZVO, the effective
spin value is close to $\mathit{S}$ = 1/2 whereas in ZVO the effective
spin is close to $\mathit{S}$ = 1. As quantum fluctuations are more
prominent in low spin systems, quantum effects in LZVO might play
a dominant role compared to classical ZVO.\textbf{ }Also,\textbf{
}Zn$_{3}$V$_{3}$O$_{8}$ (ZVO) has the ratio of V$^{3+}$: V$^{4+}$=
2:1 whereas for LZVO it is 1:2. In contrast, \textcolor{black}{Li$_{2}$ZnV$_{3}$O$_{8}$}
contains only $S=1/2$ V$^{4+}$. The strength of the exchange couplings
reflected in the Curie-Weiss temperature $\theta_{\mathrm{CW}}$ is
about -370 K for ZVO while for \textcolor{black}{Li$_{2}$ZnV$_{3}$O$_{8}$}
it is about -210 K. For both ZVO and \textcolor{black}{Li$_{2}$ZnV$_{3}$O$_{8}$}
the irreversible temperature is $\mathit{T_{irr}\simeq}$ 6.0 K and
freezing temperatures ($\mathit{T_{f}}$) is around 3.75 K and 3.50
K respectively.

In this work, we report sample preparation, structural analysis, bulk
magnetic properties, heat capacity and $^{7}$Li NMR measurements
on \textcolor{black}{LZVO}. The system crystallizes in the centrosymmetric
$\mathit{Fd\bar{\mathrm{3}}m}$ space group with a site sharing between
the lithium and the vanadium atoms at the crystallographic 16c sites.
Our magnetization data show no long-range ordering down to 2 K but
we found a splitting between the zero field cooled (ZFC) and field
cooled (FC) data at 3 K in low fields. A large Curie-Weiss temperature,
$\theta_{\mathrm{CW}}=$ - 185 K indicates strong antiferromagnetic
(AFM) interactions between the magnetic ions. Our frequency dependent
ac susceptibility results indicate a spin-glass ground state. To know
the spin-dynamics of the glassy phase in detail, we have carried out
further magnetization measurements below the freezing temperature
($T_{f}$) and observed clear signature of magnetic relaxation, aging
effect and memory phenomena which are thought to be typical characteristics
of spin-glasses. In memory effects, we observed that a small heating
cycle erases its previous memory and reinitializes the relaxation
process. This type of asymmetric response favors by the hierarchical
model \citep{Sun2003,Lefloch1992}. The inferred magnetic heat capacity
has a broad maximum at $\mathit{T}$ $\sim$ 7 K but no sharp anomaly
indicative of long-range ordering (LRO) is observed. The field swept
$^{7}$Li NMR spectra down to 1.8 K indicate the presence of an NMR
line shift and gradual broadening of the spectra as $\mathit{T}$
decreases. The spin-lattice and spin-spin relaxation rates both show
anomalies at a freezing temperature of 3 K.

\section{Experimental detail}

The polycrystalline LiZn$_{2}$V$_{3}$O$_{8}$ sample was prepared
by conventional solid-state reaction techniques using high purity
starting materials. The sample was prepared in two steps. First, we
prepared the LiZnVO$_{4}$ precursor from a stoichiometric mixture
of preheated Li$_{2}$CO$_{3}$ (99.995\% pure), ZnO (99.9\% pure)
and V$_{2}$O$_{5}$ (99.99\% pure) at $\mathrm{700\mathrm{^{o}C}}$
for 15 hours in a box furnace. After that, LiZnVO$_{4}$, V$_{2}$O$_{3}$
and ZnO were mixed well (in the molar ratio 1:1:1), pelletized and sealed
in a quartz tube after flushing with argon gas. The sample was then
fired in a box furnace for 24 hours with intermediate regrinding,
pelletization and sealing. The temperatures during successive firing
were $\mathrm{600\mathrm{\mathrm{^{o}C}}}$ and $\mathrm{700\mathrm{\mathrm{^{o}C}}}$.
Powder x-ray diffraction (XRD) measurements were performed at room
temperature (RT) with Cu $K_{\alpha}$ radiation ($\lambda=1.54182\mathrm{\,\mathring{A}}$)
on a PANalytical X\textquoteright Pert PRO diffractometer. Magnetization
measurements were carried out in the temperature range $1.8-400$
K and the field range $0-70$ kOe using a Quantum Design SQUID VSM.
For the low-field magnetization measurements, the reset magnet mode
option of the SQUID VSM was used to set the field to zero. Heat capacity
measurements were performed in the temperature range $1.8-295$ K
and in the field range $0-90$ kOe using the heat capacity option
of a Quantum Design PPMS. The NMR measurements were carried out using
a phase-coherent pulse spectrometer on $^{7}$Li nuclei in a temperature
range 1.8 - 200 K. We have measured field sweep $^{7}$Li NMR spectra,
spin-spin relaxation rate ($\frac{1}{T_{\mathrm{2}}}$) and spin-lattice
relaxation rate ($\frac{1}{T_{\mathrm{1}}}$) at two different fixed
radio frequencies (rf) of about 95 MHz and 30 MHz, respectively. The
spectra were measured using a conventional spin-echo sequence ($\mathit{\frac{\pi}{2}-\tau_{echo}-\pi}$).
We have used the spin-echo pulse sequence with a variable delay time
$\tau_{D}$ to measure $\mathit{T_{\mathrm{2}}}$ and a saturation
pulse sequence ($\frac{\pi}{2}-\tau_{D}-\pi$) is used after a waiting
time or last delay of three to five times of $\mathit{T_{\mathrm{1}}}$ between
each saturation pulse to measure $\mathit{T_{\mathrm{1}}}$.

\section{Results and Discussion}
\noindent \begin{center}
\textbf{\large{}A. Crystal structure}{\large\par}
\par\end{center}

\begin{figure}[h]
\centering{}\includegraphics[scale=0.38]{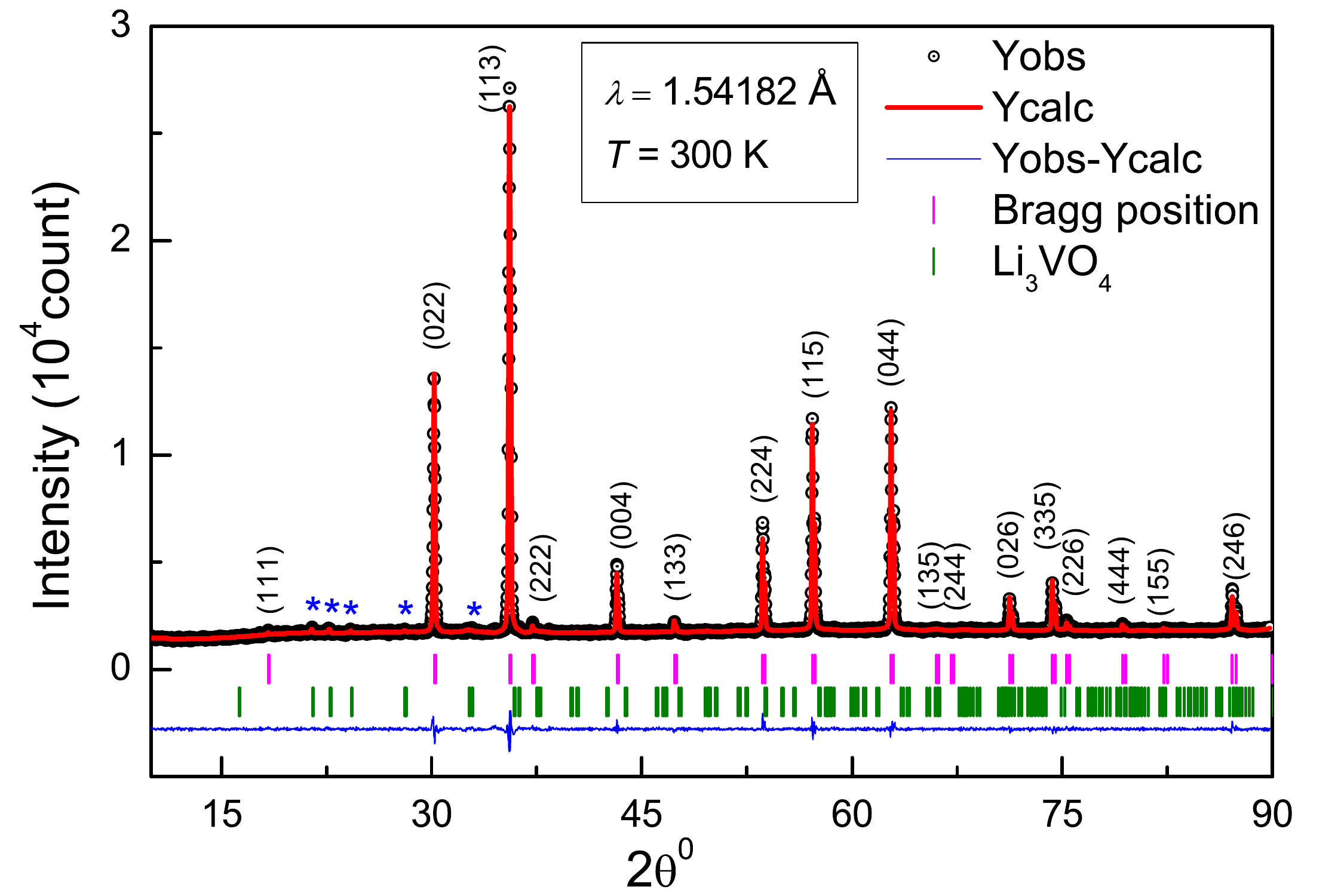}\caption{\label{fig:Two-phase-Reitveld}{\small{} Powder XRD pattern of LZVO
at 300 K and its Rietveld refinement considering }$\mathit{Fd\bar{\mathrm{3}}m}${\small{}
space group is shown along with its Bragg peak positions (pink vertical
bars) and the corresponding Miller indices (hkl). The black circles
are the observed data, the red solid line is the calculated Rietveld
pattern, the blue solid line is the residual data and the olive vertical
bars are the Bragg peak positions for the non-magnetic impurity Li$_{3}$VO$_{4}$.
The main peaks of Li$_{3}$VO$_{4}$ are indicated by blue asterisks.}}
\end{figure}

\noindent To check for phase purity, we have measured the XRD pattern
of polycrystalline LZVO. The Rietveld refinement of LZVO (shown in
Fig. \ref{fig:Two-phase-Reitveld}) revealed that it crystallizes
in the centrosymmetric cubic spinel $\mathit{Fd\bar{\mathrm{3}}m}$
(227) space group. The two phase refinement of LZVO indicates the
presence of non-magnetic impurity phase Li$_{3}$VO$_{4}$ (less than
2\%) together with the main phase. The atomic positions obtained after
Rietveld refinement (using Fullprof suite \citep{Rodriguez-Carvajal1993})
are given in Table \ref{tab:Atomic-Positions-of LZVO}. After refinement,
we obtained the lattice parameters as\textbf{ }: $\mathit{a}$ = $\mathit{b}$
= $\mathit{c}$ = 8.364 Å, ${\normalcolor \alpha}$ = ${\normalcolor \beta}$
= ${\normalcolor \gamma}=$ ${\normalcolor 90^{0}}$. The goodness
of the Rietveld refinement is inferred from the following parameters\textbf{
}${\normalcolor \chi^{2}}$: 1.37; $\mathrm{{\normalcolor R}}_{\mathbf{\mathrm{p}}}$:
21.8\%; ${\normalcolor \mathrm{R}}_{{\normalcolor \mathrm{wp}}}$:
10.7\%; ${\normalcolor \mathrm{R}}_{{\normalcolor \mathrm{exp}}}$:
9.14\%.

\noindent 
\begin{figure}[h]
\centering{}\includegraphics[scale=0.4]{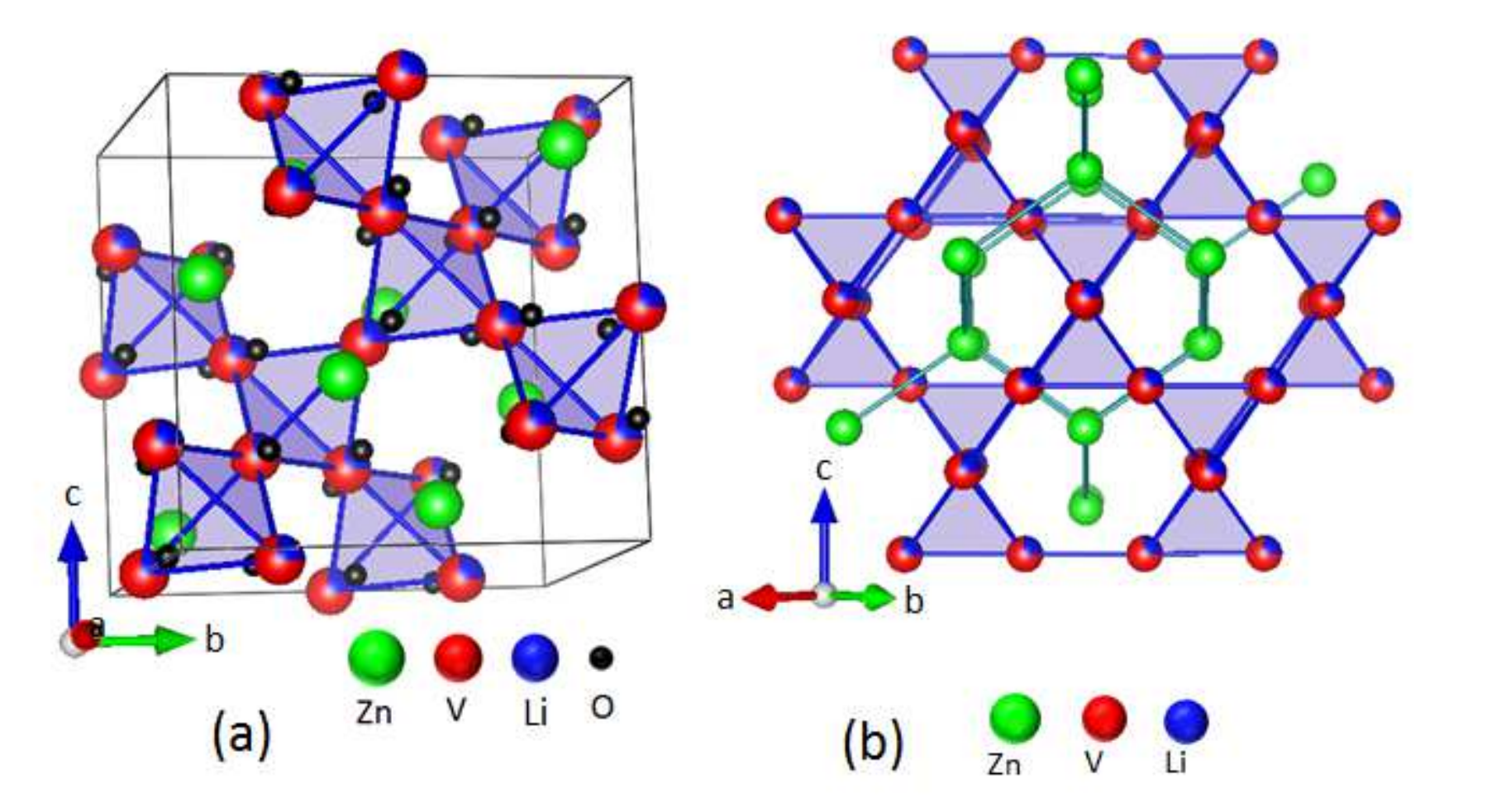}\caption{\label{fig:Structure pf LZVO}{\small{}(a) One unit cell of LZVO.
(b) Corner shared tetrahedral network of magnetic vanadium atoms and
corner shared diamond structure of Zn$^{2+}$ ions.}}
\end{figure}
Note that from Rietveld refinement we can rewrite the chemical formula
of LiZn$_{2}$V$_{3}$O$_{8}$ as {[}Zn$_{1.0}$(Li$_{0.241}$V$_{0.759}$)$_{2}$O$_{4}${]}$_{2}$
like the cubic spinel general formula AB$_{2}$O$_{4}$. One unit
cell is shown in Fig. \ref{fig:Structure pf LZVO}(a). Here the Zn$^{2+}$
ions are at the tetrahedral A-sites and connected to other Zn$^{2+}$
ions like in a diamond structure and the octahedral B-sites are statistically
occupied by Li$^{1+}$ and vanadium ions (both V$^{3+}$ and V$^{4+}$)
in a 1:3 ratio. The V-V bond distance is 2.95 Å. The vanadium occupies
the octahedral B-site and among themselves they form a corner shared
tetrahedral network like in the pyrochlore lattice which is geometrically
frustrated. Random or statistical distribution between the Li$^{1+}$,
V$^{3+}$ and V$^{4+}$ ions dilutes the corner-shared tetrahedral
magnetic network. This dilution works as an additional source of disorder
and forces the system to relieve the frustration and a spin-glass
or frozen state might emerge. Fig. \ref{fig:Structure pf LZVO}(a)
and (b) shows (Li/V)$_{4}$ corner shared tetrahedra in 3D. 
\begin{table}[h]
\centering{}\caption{\label{tab:Atomic-Positions-of LZVO}{\small{}Atomic positions in
LiZn$_{2}$V$_{3}$O$_{8}$ after Rietveld refinement of powder XRD
at room temperature.}}
\vspace{0.5cm}
\begin{tabular}{cccccc}
\hline 
Atom & Wyckoff position & x/a & y/b & z/c & Occupancy\tabularnewline
\hline 
\hline 
Zn & 8b & 0.375 & 0.375 & 0.375 & 1.000\tabularnewline
Li & 16c & 0.000 & 0.000 & 0.000 & 0.241\tabularnewline
V & 16c & 0.000 & 0.000 & 0.000 & 0.759\tabularnewline
O & 32e & 0.239 & 0.0.239 & 0.239 & 1.000\tabularnewline
\hline 
\end{tabular}
\end{table}

\begin{center}
\textbf{\large{}B. DC susceptibility} 
\par\end{center}

\begin{figure}[h]
\centering{}\includegraphics[scale=0.35]{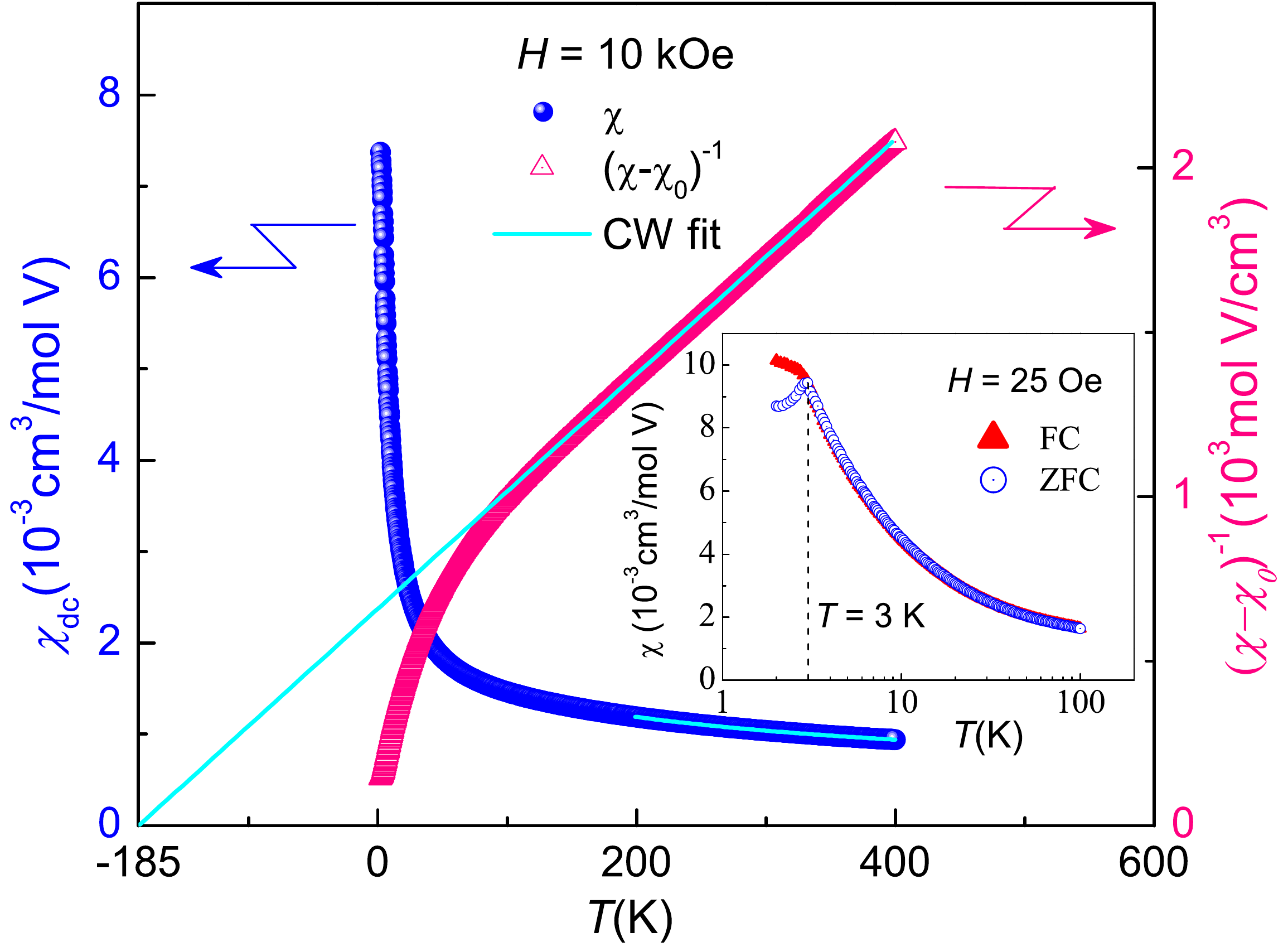}\caption{\label{fig:Chi of LZVO}The dc susceptibility $\chi_{dc}(T)$ (on
left $\mathit{y}$-axis) and the inverse susceptibility free from
$\mathit{T}$-independent $\chi$ (on right $\mathit{y}$-axis) of
LZVO in $\mathit{H}$ = 10\,kOe are depicted as a function of temperature.
The solid lines show the Curie-Weiss fit. In the inset, the bifurcation
between zero field cooled (ZFC) and field cooled (FC) data below $T_{f}$
= 3 K is observed in an applied field of 25 Oe.}
\end{figure}
The magnetization ($M$) of LZVO was measured as a function of temperature
($T$) at various applied magnetic fields ($H$). The dc susceptibility
($\mathit{\chi}_{\mathrm{dc}}(T)$ = $\mathit{\frac{M}{H}}$) is plotted
as a function of temperature in an applied field $H$ = 10 kOe in
Fig. \ref{fig:Chi of LZVO}. There is no long-range or short-range
ordering down to 2 K. The $\mathit{\chi}_{\mathrm{dc}}$($\mathit{T}$)
in $H$ = 10 kOe is paramagnetic and follows a Curie-Weiss law. The
zero field cooled (ZFC) and field cooled (FC) data in $\mathit{H}$
= 10 kOe show no difference down to 2 K. From a Curie-Weiss fit of
$\chi_{\mathrm{dc}}(T)$ using the equation: $\chi(T)=\chi_{0}+\frac{C}{(T-\theta_{\mathrm{CW}})}$
in the range of (200 - 400) K, we obtained the temperature independent
susceptibility $\chi_{0}$ = $4.62\times10^{-4}$ (cm$^{3}$/mol V),
the Curie constant $C$ = 0.28 Kcm$^{3}$/mol V and the Curie-Weiss
(CW) temperature, $\theta_{\mathrm{CW}}$ = -185 K. The effective
moment $\mu_{\mathrm{eff}}$ is then about 1.50 $\mu_{\mathrm{B}}$
which is less than that of the $\mu_{\mathrm{eff}}$ = 1.73 $\mu_{\mathrm{B}}$
for a spin $\mathit{S}$ = $\frac{1}{2}$ system. From the stoichiometry
of LZVO, one might have expected two V$^{4+}$ ions and one V$^{3+}$
ion per formula unit giving a Curie constant $\mathit{C}$ = 0.583
cm$^{3}$K/mol V. However, we inferred a much smaller value. This
result is similar to the homologous cluster magnet LiZn$_{2}$Mo$_{3}$O$_{8}$
\citep{Sheckelton2012} where the Curie constant ($\mathit{C}$ =
0.24 Kcm$^{3}$/mol f.u.) was less than the expected value. In metallic
LiV$_{2}$O$_{4}$, the Curie constant was found to be that corresponding
to one $\mathit{S}$ = $\frac{1}{2}$ moment per vanadium whereas
one has, on an average, 1.5 electrons per vanadium. It is possible
that here as well, the moment is quenched to some extent due to frustration.
The high CW temperature suggests strong antiferromagnetic interactions
between the magnetic vanadium atoms in the sample. In the inset of
Fig. \ref{fig:Chi of LZVO}, the susceptibility, in a low field of
25 Oe, in ZFC (open circles) and FC (closed triangles) mode shows
bifurcation below $T_{f}$ $\simeq$ 3 K. This bifurcation suggests
a spin-glass (SG) ground state of the system at low temperature. The
inferred frustration parameter \citep{Greedan2001} ($f=\frac{\mid\theta_{\mathrm{cw}}\mid}{T_{\mathrm{N}}}$ $\approx62$)
puts the LZVO system in the highly frustrated category.
\begin{center}
\textbf{\large{}C. AC susceptibility}{\large\par}
\par\end{center}

\begin{figure}[h]
\centering{}\includegraphics[scale=0.32]{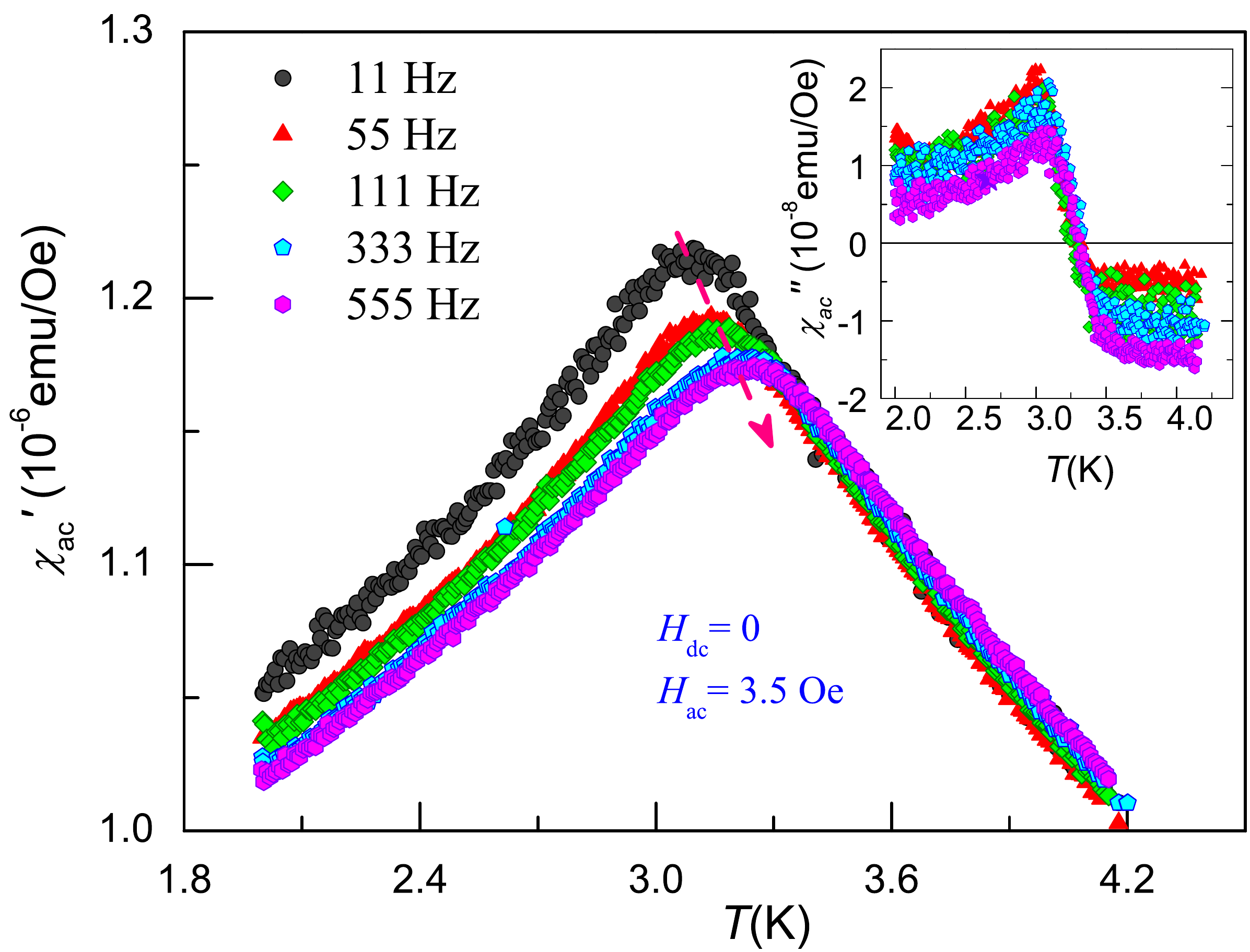}\caption{\label{fig:ac susceptibility of LZVO}{\small{} The temperature dependence
of the in phase $\chi_{\mathrm{ac}}^{'}$ (in main figure) and the
out of phase $\chi_{\mathrm{ac}}^{''}$ component (in the inset) of
ac susceptibility ($\chi_{\mathrm{ac}}$) at different frequencies
are shown.}}
\end{figure}

To further understand the nature of the SG states of LZVO, the ac
magnetic susceptibility was measured in a fixed ac field of $H_{ac}$
= 3.5 Oe with \textit{T}- range (2\ensuremath{\le} \textit{T} \ensuremath{\le}
4.2 K) and frequency range 1 \ensuremath{\le} $\nu$ \ensuremath{\le}
1000 Hz. The temperature variation of the in-phase component of the
ac susceptibility, $\chi_{\mathrm{ac}}^{'}(T)$, as shown in Fig.
\ref{fig:ac susceptibility of LZVO}, shows a cusp like behavior and
a peak at $T_{f}$ = 3.0 K for $\nu$ = 11 Hz. The peak position shifts
towards higher temperatures as the frequency $\mathit{\nu}$ increases
from 11 Hz to 555 Hz. So the frequency dependence of $\chi_{\mathrm{ac}}^{'}(T)$
is evident with a downward shift of the anomaly with increasing $\nu$
and is consistent with SG systems. Even the out of phase component
of ac susceptibility $\chi_{\mathrm{ac}}^{''}(T)$ shows a peak at
$T_{f}$ which barely shifts with frequency variation (shown in the
inset of Fig. \ref{fig:ac susceptibility of LZVO}). The value of
$\chi_{\mathrm{ac}}^{''}(T)$ is non-zero positive below $T_{f}$
and is negative above $T_{f}$. This indicates the SG state is associated
with frustrated magnets.
Such behavior is distinctive of the SG states and allows us to categorize
the SG systems from the disordered antiferromagnetic (AFM) systems.
In the disordered AFM systems, the value of $\chi_{\mathrm{ac}}^{''}(T)$
is constant and remains zero below the transition temperature \citep{Malinowski2011,Mulder1982,Suellow1997}.
All these features confirm the formation of a SG ground state of
LZVO. 

The frequency dependence of $T_{f}$ is often quantified in terms
of the relative shift of the spin freezing temperature, defined as
$\delta T_{f}$ = {[} $\Delta T_{f}$ / $T_{f}$ $\mathit{\mathrm{\Delta}}$log$_{10}$($\mathit{\nu}$){]}
\citep{Mahendiran2003}. This relative shift paramenter $\delta T_{f}$
which is also known as Mydosh parameter \citep{Mydosh2015} is used
to identify different SG systems. The calculated value is $\delta T_{f}$
$\simeq$ 0.039 for our LZVO. This value of $\mathit{\delta T_{f}}$
indicates that the sensitivity to the frequency of LZVO is larger
by an order of magnitude than that for canonical SG systems such as
CuMn ($\mathit{\delta T_{f}}$ = 0.005) \citep{Mydosh1993} and AuMn
($\mathit{\delta T_{\mathrm{f}}}$ = 0.0045) \citep{Mulder1982}.
Indeed, it is intermediate between the values for canonical SG systems
and superparamagnets ($\delta T_{f}$ $\simeq$ 0.28). Also the present
value of $\delta T_{f}$ is close to that of the shape memory alloys
showing re-entrant spin-glass (RSG) behavior \citep{Chatterjee2009},
0.037 seen in metallic glasses \citep{Luo2008} and 0.095 reported
in LaCo$_{0.5}$Ni$_{0.5}$O$_{3}$ \citep{Viswanathan2009}.
\begin{figure}[h]
\centering{}\includegraphics[scale=0.45]{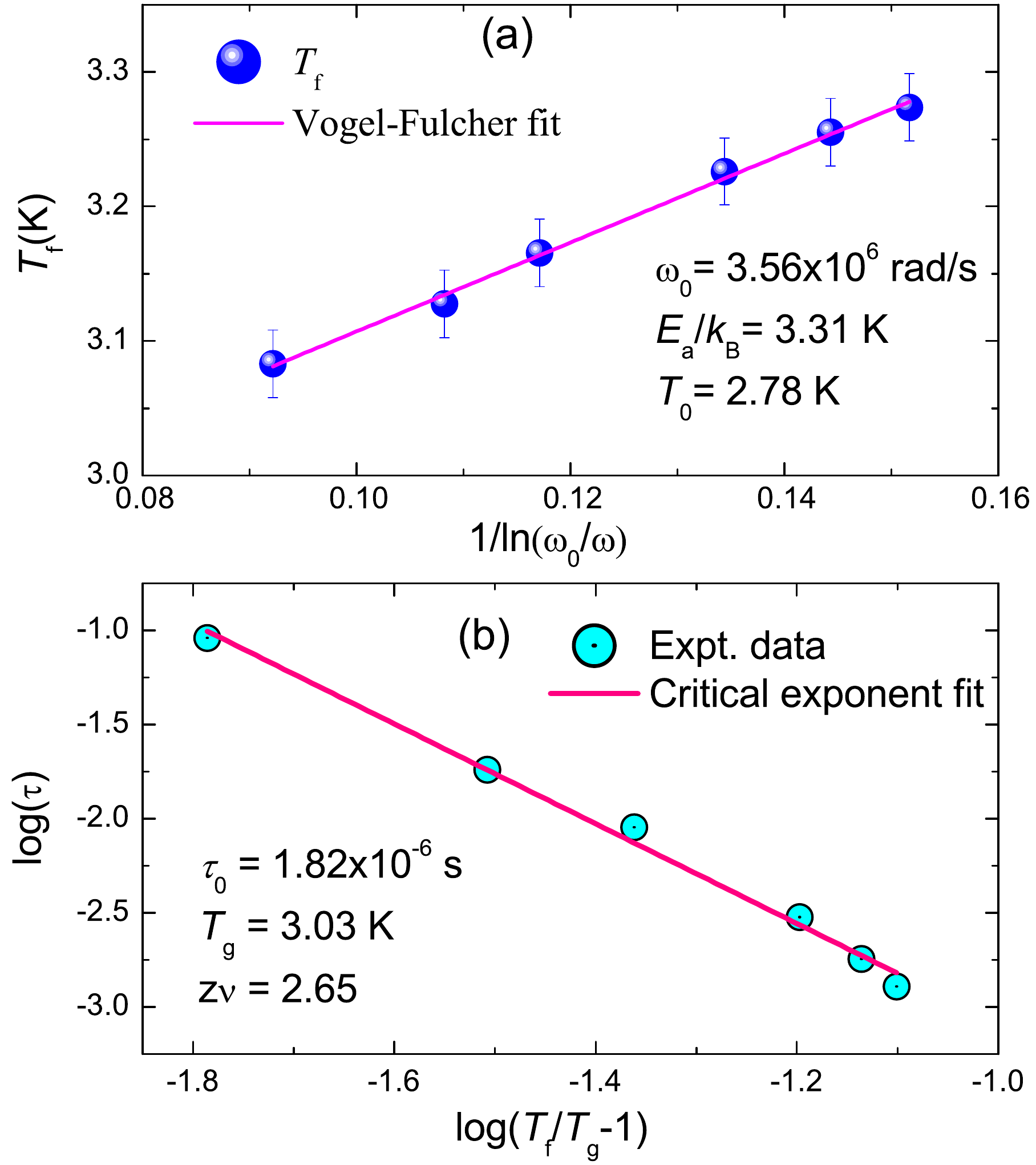}\caption{\label{fig:VF+ZV fit LZVO} {\small{}Frequency dependence of the freezing
temperature }$T_{f}${\small{} along with (a) the fit to the Vogel-Fulcher
law. (b) the fit to the critical slowing down formula (see text).}}
\end{figure}

To know the dynamic properties of the spins, we have fitted $T_{f}$
with the empirical Vogel\textendash Fulcher law: $\omega=\omega_{0}\thinspace\thinspace exp[-\mathit{\frac{E_{a}}{k_{B}\mathrm{(}T_{f}-T_{0}\mathrm{)}}}]$
in Fig. \ref{fig:VF+ZV fit LZVO}(a). Here $\omega_{0}$ is the characteristic
angular frequency, $\omega$ is the angular frequency ($\omega$ =
2$\mathit{\pi\nu}$), $\mathit{E}$$_{a}$ and $\mathit{T_{\mathrm{0}}}$
are the activation energy and Vogel\textendash Fulcher temperature,
respectively. The best fit, shown in Fig. \ref{fig:VF+ZV fit LZVO}(a),
is obtained for $\omega_{0}$ $\approx$ 3.56 \texttimes{} 10$^{6}$
Hz, $\mathit{E_{a}/k_{B}}$ $\approx$ 3.31$\pm0.10$ K, and $\mathit{T_{\mathrm{0}}}$
$\approx$ 2.78$\pm0.01$ K. As the measured frequency range is limited
(only up to 1 kHz), the error bar will be high in the derived parameters
from such a fit. The Vogel-Fulcher fit of the variations of the freezing
temperature ($T_{f}$) with frequency suggests short-range Ising SG
behavior \citep{Fisher1986}. The value of $\omega_{0}$ obtained
from the fitting is less than that of conventional SG systems, which
is expected to be about $10^{13}$ rad/s. Such a low value of $\omega_{0}$
is associated with the re-entrant spin-glass (RSG) systems like Ni$_{2}$Mn$_{1.36}$Sn$_{0.64}$
\citep{Chatterjee2009} and cluster spin-glass (CSG) Zn$_{3}$V$_{3}$O$_{8}$
\citep{Chakrabarty2014}. Thus we might suggest that the spin glass
state in LZVO is not atomic in origin; rather it is related to flipping
of large number of spins simultaneously or in a collective manner
which leads to the formation of CSG states.

On further analysis, the $T_{f}$ is found to obey the critical slowing
down dynamics (see Fig. \ref{fig:VF+ZV fit LZVO}(b)) governed by
the equation: $\tau=\tau_{0}(\frac{T_{f}}{T_{\mathrm{g}}}-1)^{-zv}$,
where $\tau_{0}$ is spin flipping relaxation time of the fluctuating
entities, $\mathit{zv}$ is the dynamic exponent and $T_{\mathrm{g}}$
is the static freezing temperature \citep{Dho2002}. We found the
best fit with $T_{\mathrm{g}}$ = 3.03$\pm0.01$ K, $\tau_{0}$ $\approx$
1.82 $\times$10$^{-7}$ s and $\mathit{zv}$ $\approx$ 2.65. For
the conventional SG systems, the value of $\tau_{0}$ ranges within
10$^{-10}$ to 10$^{-13}$ s and $\mathit{zv}$ in the range (4 -
13) \citep{Luo2008}. The fact that the present value of $\tau_{0}$,
is higher than that of the conventional SG systems, suggests that
in LZVO, the spin-dynamics develops at a slower rate due to the presence
of randomly magnetized interacting clusters, instead of individual
spin randomness. Similar values have also been reported in SG systems
such as Heusler alloys, LaCo$_{0.5}$Ni$_{0.5}$O$_{3}$, pyrochlore
molybdates etc. \citep{Chatterjee2009,Hanasaki2007,Viswanathan2009}.
The value of $\mathit{z\nu}$ falls in the range of other typical
CSG systems.
\begin{center}
\textbf{\large{}D. Aging effect and relaxation}{\large\par}
\par\end{center}

\begin{figure}[H]
\centering{}\includegraphics[scale=0.45]{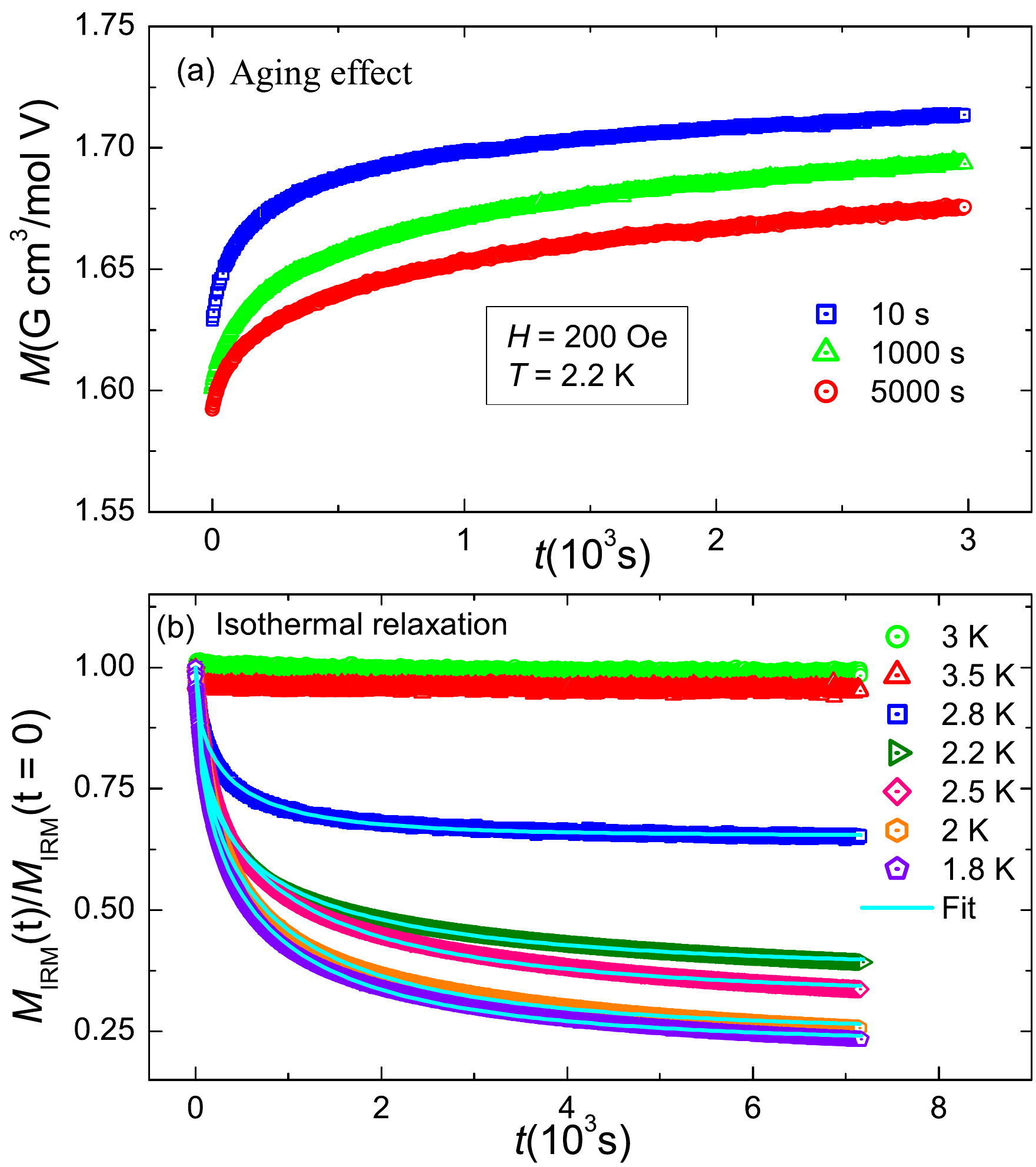}\caption{\label{fig:Aging-effect-LZVO}(a) Aging effect shows the growth of
the magnetization as a function of time with three different waiting
times and (b) Isothermal remanent magnetic relaxations (normalized
with respect to the moment at $\mathit{t}$ = 0) with their fitting
(described in text) are shown as a function of time at several temperatures.}
\end{figure}

Aging effect is a characteristic signature of any glassy system \citep{Mydosh1993,Nam2000}.
Aging indicates that the response of a system becomes slower and slower
with time. Here, aging effect has been studied with time evolution
of zero field cooled (ZFC) magnetization. The Fig. \ref{fig:Aging-effect-LZVO}(a)
depicts the growth of the magnetization data as a function of time,
in the frozen state. The sample was cooled from RT to 2.2 K in the
ZFC mode and then the system was allowed to age for a waiting time
t$_{w}$. Subsequently, a field of 200 Oe was applied and the magnetization
was then recorded as a function of time. We have measured aging effect
for three different waiting times 10 s, 1000 s and 5000 s. It is clear
that the magnetization growth is slower for larger waiting times,
which indicates the metastability of the low temperature magnetic
state. We also measured the isothermal remanent magnetization ($M_{\mathrm{IRM}}$)
of LZVO to explore the metastable behavior of the SG state around
the SG transition temperature. To measure the $M_{\mathrm{IRM}}$,
first we cooled the sample in the ZFC mode from RT to the desired
temperature mainly below the transition point, then we applied a field
of 500 Oe and we switched off the field after 300 s and allowed the
system to relax. During relaxation, we recorded the magnetization
as a function of time for two hours. Fig. \ref{fig:Aging-effect-LZVO}(b)
shows seven relaxation curves which are normalized to the magnetization
before making the field zero, $M_{\mathrm{IRM}}(t)/M_{\mathrm{IRM}}(0)$
at seven different temperatures. The time dependence of $\mathit{M_{\mathrm{IRM}}\mathrm{(}t\mathrm{)}}$
is well fitted with the stretched exponential given as in equation
$M_{\mathrm{t}}(H)=M_{0}(H)+[M_{\infty}(H)-M_{0}(H)][1-exp\{-(t/\tau)^{\alpha}\}]$.
Here, $\mathit{M_{\mathrm{0}}}$ and $\mathit{M_{\infty}}$ are magnetizations
at t $\rightarrow$ 0 and t $\rightarrow$ $\infty$, $\tau$ is the
characteristic relaxation time and $\alpha$ is the stretching exponent,
which ranges between 0 and 1. Depending on the value of $\alpha$,
one can have an estimation of the distribution of energy barriers
present in the frozen state. The obtained best fit parameters for
each isotherm are listed in Table \ref{tab:relaxation time}. We have
found that the values of $\alpha$ for LZVO, varies within 0.45 to
0.55, is consistent with earlier reported glassy systems \citep{Mydosh2015,Bhattacharyya2011,Chakrabarty2014}.
Also $\alpha$ < 1 indicates the anisotropic nature of energy barries
i.e. the presence of metastable states in our LZVO system. From the
value of $\tau$, it is evident that the decay is faster as the temperature
near  $T_{f}$ $\simeq$ 3 K. This also signifies that LZVO system goes
to a metastable and irreversible state below $T_{f}$ on application
of field. As expected, when $\mathit{T}$ $\geq$ $T_{f}$, $M_{\mathrm{IRM}}$($\mathit{t}$)
is independent of time.

\begin{table}[h]

\caption{\label{tab:relaxation time}The best fit (see text) result of isothermal
remanent magnetic relaxation of LZVO.}
\smallskip{}

\begin{centering}
\begin{tabular}{cccc}
\hline 
$\mathit{T}$(K) & $\frac{M_{\infty}}{M_{0}}$ & Stretching exponent $\alpha$ & Relaxation time $\tau$(s)\tabularnewline
\hline 
\hline 
1.8 & 0.223 & 0.53 & 580\tabularnewline
2.0 & 0.248 & 0.55 & 649\tabularnewline
2.2 & 0.372 & 0.46 & 593\tabularnewline
2.5 & 0.318 & 0.52 & 737\tabularnewline
2.8 & 0.653 & 0.54 & 323\tabularnewline
\hline 
\end{tabular}
\par\end{centering}
\end{table}

\begin{figure}[h]
\centering{}\includegraphics[scale=0.35]{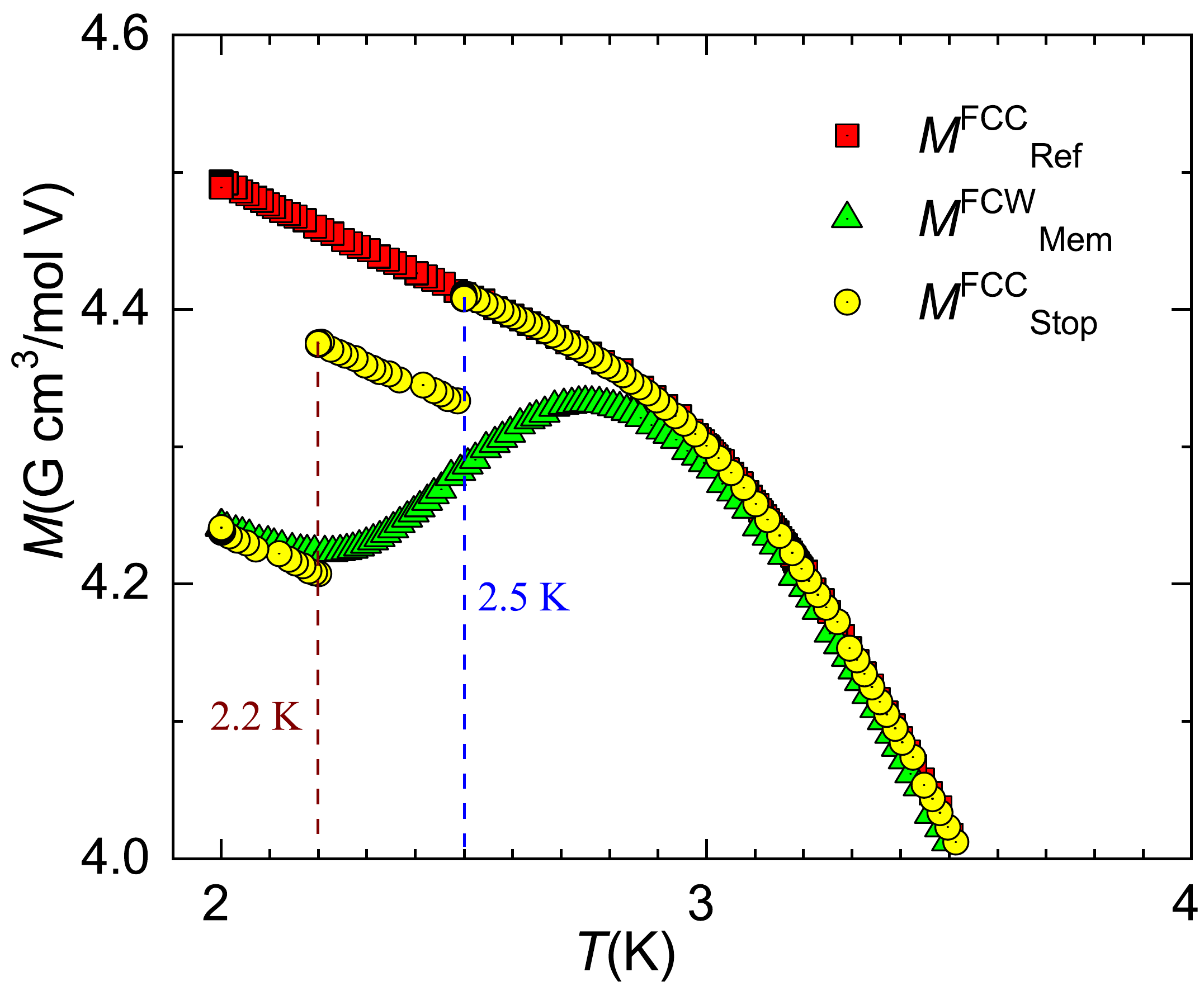}\caption{\label{fig:The-memory-effect LZVO}{\small{}The memory effect in LZVO
as a function of temperature is observed in FC magnetization mode
at $\mathit{H}$ = 500 Oe. The $M_{\mathrm{Stop}}^{\mathrm{FCC}}$
curve was obtained during cooling the sample with intermediate stops
of 3 hours duration each at 2.5 K and 2.2 K. The $M_{\mathrm{Mem}}^{\mathrm{FCW}}$
and $M_{\mathrm{Ref}}^{\mathrm{FCC}}$curves were measured during
continuous heating and cooling of the sample respectively at $\mathit{H}$
= 500 Oe.}}
\end{figure}

\begin{center}
\textbf{\large{}E. Memory effect}{\large\par}
\par\end{center}

Fig. \ref{fig:The-memory-effect LZVO} shows a memory effect which
is measured in the FC magnetization mode using the protocol introduced
by Sun $\mathit{et}$ $\mathit{al.}$ \citep{Sun2003}. In an applied
field of \textit{H} = 500 Oe, we have recorded the magnetization of
LZVO as a function of temperature from 100 K down to 2 K with a cooling
rate of 1 K/min. Below $T_{f}$, We interrupted the cooling process
twice at 2.5 K and 2.2 K for a waiting time $t_{\mathrm{w}}$ = 3
hr each. We switched off the field during $t_{\mathrm{w}}$ and let
the system to relax. We resumed the FC process after each stop and
wait period. The stops at 2.5 K and 2.2 K are obvious as step-like
features obtained in the $M_{\mathrm{Stop}}^{\mathrm{FCC}}$ curve
shown in Fig. \ref{fig:The-memory-effect LZVO}. After reaching  2 K, we heated the sample continuously in the same magnetic
field and recorded the magnetization data simultaneously. The magnetization
obtained in this way, referred to as $M_{\mathrm{Mem}}^{\mathrm{FCW}}$,
exhibits a slight inflexion at 2.5 K and a pronounced minimum
at 2.2 K. This indicates that somehow the system has its previous
behavior imprinted as a memory during the cooling process. Similar
behavior has been noticed earlier in intermetallic compounds such
as GdCu \citep{Bhattacharyya2011}, Nd$_{5}$Ge$_{3}$ \citep{Maji2011}
and in super-spin-glass nanoparticle systems \citep{Sun2003,Sasaki2005}.
This is nothing but the typical aspect of SG systems. The inflexion point
or dip at 2.5 K in the $M_{\mathrm{Mem}}^{\mathrm{FCW}}$ curve is
weak because at 2.5 K the system is not much below the freezing temperature
($\mathit{i.e.}$ $T_{f}$ = 3.0 K at $\mathit{H}$ = 500 Oe). A reference
curve ($M_{\mathrm{Ref}}^{\mathrm{FCW}}$) was also measured by cooling
the sample continuously at $\mathit{H}$ = 500 Oe. It is worth noting
that there was no memory effect when we wait at a temperature
above $T_{f}$.

\begin{figure}[h]
\centering{}\includegraphics[scale=0.35]{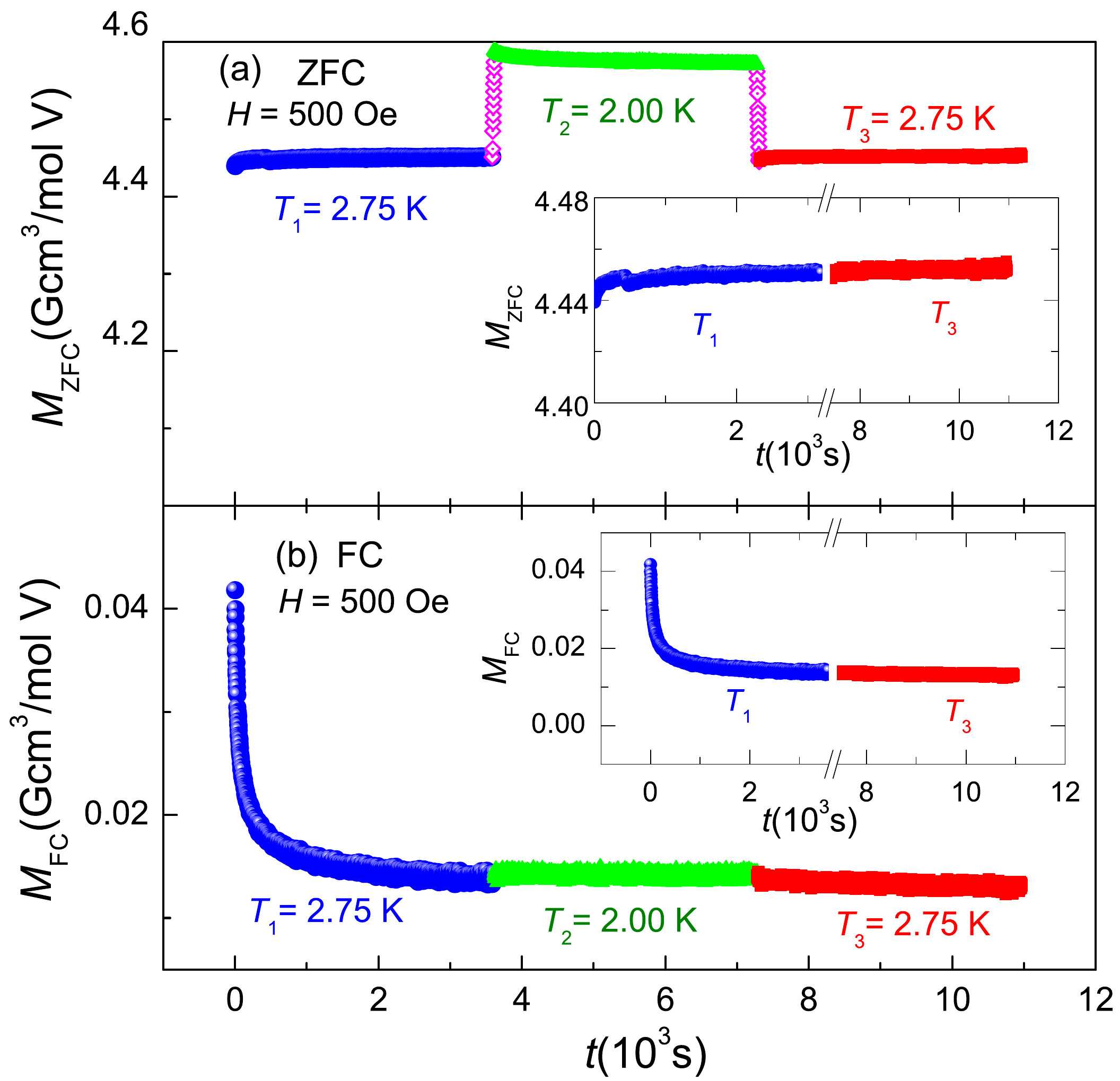}\caption{\label{fig:Magnetic-relaxation-in LZVO}{\small{}The magnetic relaxation
in LZVO at 2.75 K with a negative heating cycle at $\mathit{H}$ =
500 Oe for (a) the ZFC and (b) the FC methods. The insets show the
relaxation data during $T_{1}$ and $T_{3}$ which are merged in a seamless manner.}}
\end{figure}

\textit{Negative heating cycle:} The memory effect was further clarified
by the ZFC and FC method with a negative heating cycling as shown
in Fig. \ref{fig:Magnetic-relaxation-in LZVO}. In the ZFC mode, we
cooled the sample below the spin freezing temperature ($T_{f}$) in
zero field from the paramagnetic state to the measuring temperature
$\mathrm{\mathit{T}}_{1}$ = 2.75 K. Then we recorded the magnetization
as a function of time ($\mathit{t}$) for 1 hr in an applied field
of 500 Oe. The magnetization increases logarithmically with $\mathit{t}$.
After that, we cooled the sample temperature to a lower value $\mathit{\mathrm{\mathit{T_{\mathrm{2}}}}}$
= 2.0 K in the same field and then we have measured the magnetization
for a time $\mathrm{\mathit{t}}_{2}$ = 1 hr. Finally, we restored
the sample temperature to $\mathrm{\mathit{T}}_{1}$ = 2.75 K and
measured the magnetization for a period of $\mathrm{\mathit{t}}_{3}$
= 1 hr. The relaxation curve obtained in this way is shown in Fig.
\ref{fig:Magnetic-relaxation-in LZVO}(a). In each relaxation process,
the magnetization increases logarithmically with $\mathit{t}$ and the
system remembers its previous value it reached before the temporary
cooling started. This determines that the negative $\mathit{T}$ -
cycle does not erase the memory in ZFC mode. For the FC mode, we field
cooled the sample to $\mathit{T_{\mathrm{1}}}$ = 2.75 K first in
a field of 500 Oe. Once we reached the measuring temperature, we switched
off the field and subsequently measured the magnetization as a function
of time. Fig. \ref{fig:Magnetic-relaxation-in LZVO}(b) shows the
FC relaxation process where the magnetization decays exponentially
with $\mathit{t}$. Similarly, the FC method also preserves the state
of the system even after a temperature cooling. Inset of Fig. \ref{fig:Magnetic-relaxation-in LZVO}
(a) and (b) shows that in both ZFC and FC methods, the magnetic relaxation
during $\mathrm{\mathit{t}}_{3}$ is a continuation of the relaxation
curve during $\mathrm{\mathit{t}}_{1}$. This is a simple demonstration
of the memory effect.
\begin{figure}[h]
\centering{}\includegraphics[scale=0.35]{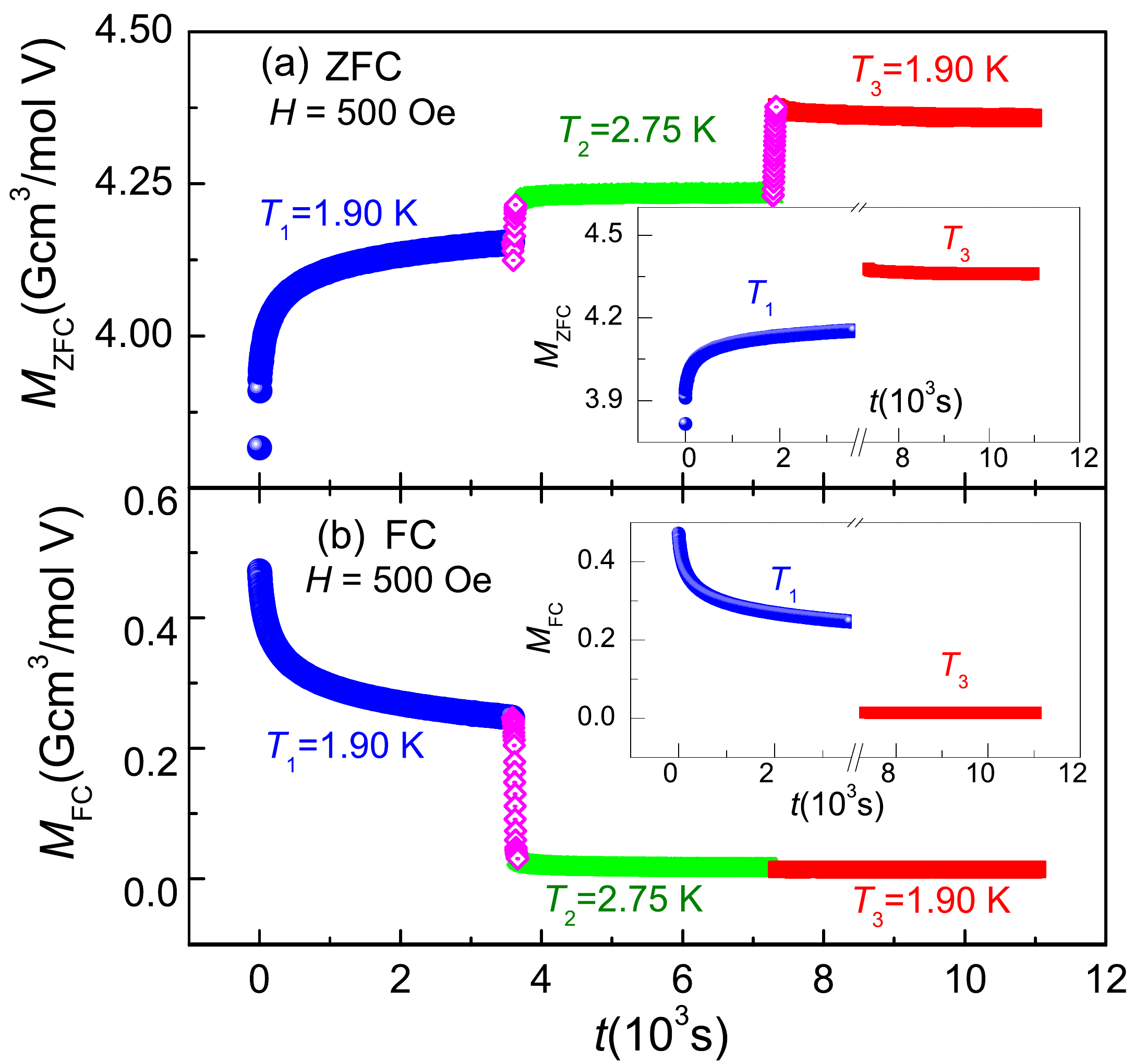}\caption{\label{fig:positive heat cycle LZVO}{\small{}The magnetic relaxation
in LZVO at 1.9 K with a positive heating cycle in $\mathit{H}$ =
500 Oe (a) the ZFC and (b) the FC methods. Insets show the relaxation
data during $T_{1}$ and $T_{3}$ which are not merged.}}
\end{figure}
\begin{figure}[h]
\begin{centering}
\includegraphics[scale=0.35]{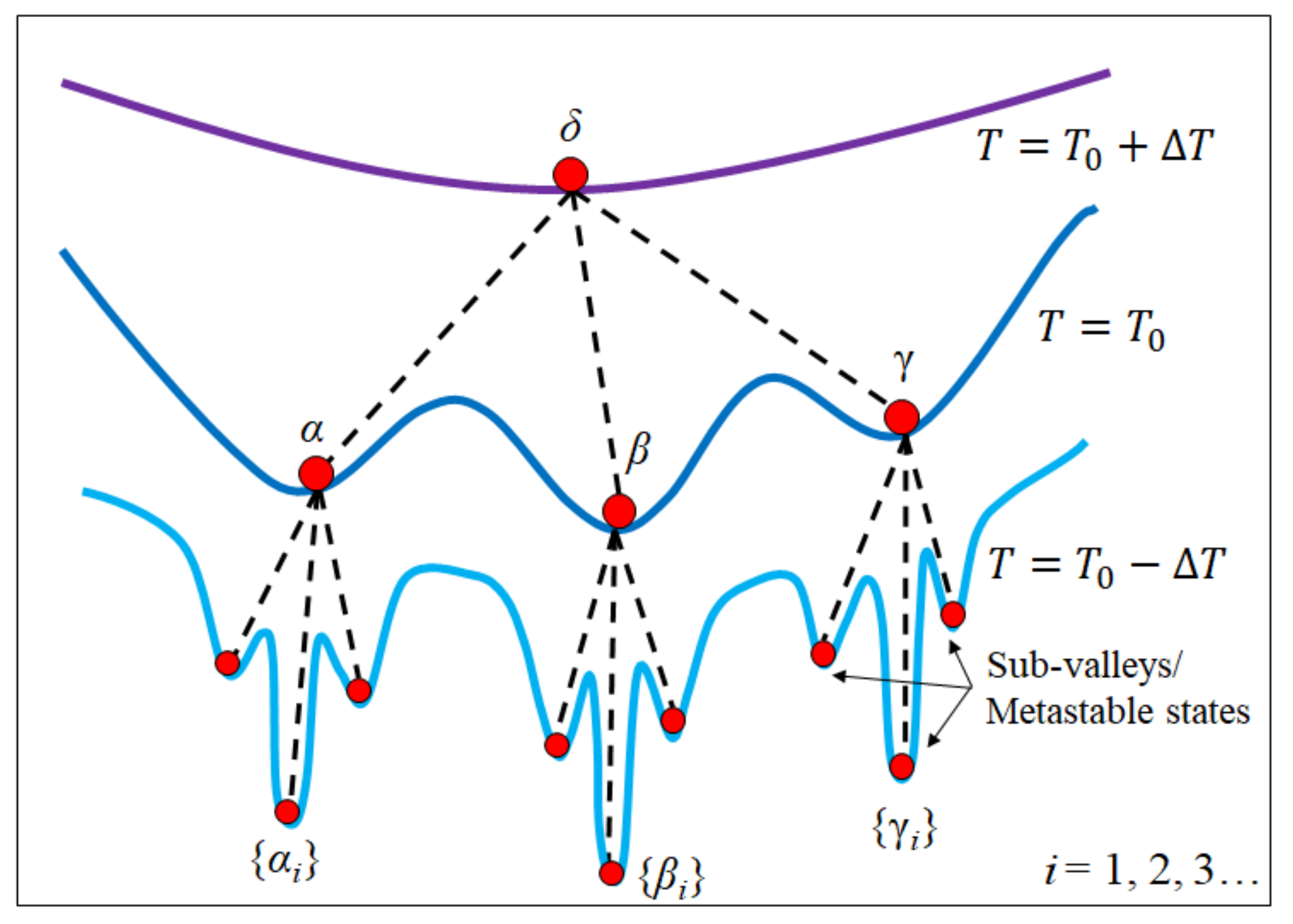}\caption{\label{fig:Hierarchical model}Schematic diagram of the hierarchical
model for spin-glass states. The different curves represent the free
energy at that particular temperatures. The metastable states of the
spin-glass are depicted by the lobs at $\mathit{T}$ = $\mathit{T_{\mathrm{0}}}$ - $\Delta$$\mathrm{\mathit{T}}$.}
\par\end{centering}
\end{figure}

\textit{Positive heating cycle:} The droplet model \citep{Fisher1988,Fisher1988a}
of SG systems supports a symmetric behavior in the magnetic relaxation
concerning either heating or cooling cycles whereas the hierarchical
model \citep{Lefloch1992,Sun2003} predicts a asymmetric response.
To compare the response of intermittent heating and cooling cycles,
we performed the relaxation experiment with a positive heating cycle
also. The results are shown in Fig. \ref{fig:positive heat cycle LZVO}(a)
and (b). One might notice that a positive temperature cycle erases
the earlier memory and re-initializes the relaxation process in both
ZFC and FC mode. This confirms that the magnetic response of the our
system irrespective of heating or cooling cycle is not symmetric.
So this study supports the hierarchical model proposed for SG systems.
Fig. \ref{fig:Hierarchical model} depicts the schematic diagram of
the hierarchical model. At a given temperature $\mathit{T_{\mathrm{0}}}$,
there exists a multi-valley \{$\alpha,$$\beta,\gamma$\} free-energy
surface for a frustrated system. When we cool the system from $\mathit{T_{\mathrm{0}}}$
to $\mathit{\mathit{T_{\mathrm{0}}}}$\textminus \textgreek{D}$\mathit{T}$,
each valley splits into many sub-valleys \{$\alpha_{i}$\}, $\{\beta_{i}\},\{\gamma_{i}$\}.
When \textgreek{D}$\mathit{T}$ is large, the energy gaps between
the primary valleys become high and the system fails to overcome this
energy barrier within a finite waiting time $t_{2}$. Therefore, the
relaxation occurs only within the sub-valleys or metastable states.
When the temperature of the system is restored to its initial value
$\mathit{T_{\mathrm{0}}}$, then the sub-valleys merge back to the
original free energy surface and relaxation at $\mathit{\mathit{T_{\mathrm{0}}}}$
resumes without being perturbed by the intermediate relaxations at
$\mathit{T_{\mathrm{0}}}$\textminus{} \textgreek{D}$\mathit{T}$.
But, if we increased the system temperature from $\mathit{T_{\mathrm{0}}}$ to
$\mathit{\mathit{T_{\mathrm{0}}}}$ + \textgreek{D}$\mathit{T}$,
then the barriers between the free energy primary valleys becomes
low or sometimes they even get merged. Then, the relaxations can easily
take place within different valleys. When the temperature is lowered
back to $\mathit{\mathit{T_{\mathrm{0}}}}$, the relative occupancy
of each energy valley does not remain the same as before even though free
energy surface goes back to the original state. Thus the state of
the system varies after a temporary heating cycle and results without
a memory effect. The behavior we found in our system, has been seen
in some other SG systems too \citep{Kundu2019,Chakrabarty2014,Bhattacharyya2011,Maji2011,Sun2003}.
\begin{figure}[h]
\centering{}\includegraphics[scale=0.38]{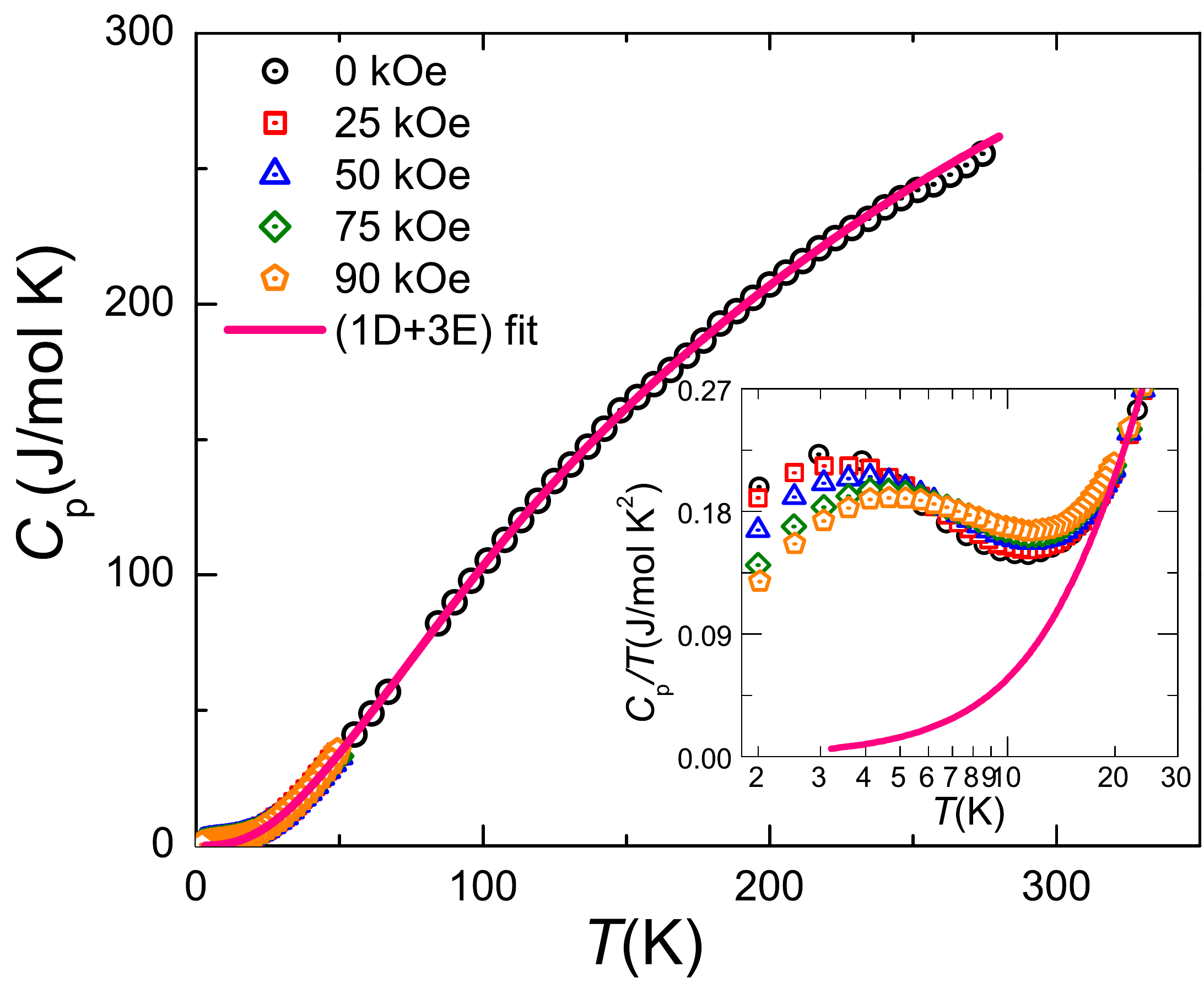}\caption{\label{fig:HC of LZVO}{\small{}The heat capacity} $C_{\mathrm{p}}(T)${\small{}
of LZVO as a function of temperature in different fields is plotted.
The pink solid line represents the (Debye + Einstein) fit (see text)
as well as the lattice heat capacity. The inset shows low temperature
anomaly of $C_{\mathrm{p}}/T$ vs. $\mathit{T}$ plot in semi-log
scale.}}
\end{figure}

\begin{center}
\textbf{\large{}F. Heat capacity}{\large\par}
\par\end{center}

We have measured the heat capacity of the LZVO sample at a constant
pressure $C_{\mathrm{p}}(T)$ by the thermal relaxation method in
different fields (0 - 90 kOe) at the temperature range (2 - 280) K
for zero field and (2 - 50) K for higher fields. The Fig. \ref{fig:HC of LZVO}
shows the temperature dependence of the specific heat at constant
pressure $C_{\mathrm{p}}(T$) of LiZn$_{2}$V$_{3}$O$_{8}$. Absence
of sharp anomaly in the $C_{\mathrm{p}}(T)$ vs. $\mathit{T}$ data
implies a lack of long-range order in LZVO and supports our dc susceptibility
data in $\mathit{H}$ = 10 kOe. Below 10 K, in the $C_{\mathrm{p}}$/$\mathit{T}$
vs. $\mathit{T}$ plot (see inset of Fig. \ref{fig:HC of LZVO}),
there is a broad hump or anomaly at low temperature. These anomalies
are field dependent and appear to be Schottky anomalies. To extract
the magnetic contribution to the heat capacity, we have to subtract
the lattice and Schottky contributions from the total heat capacity
$\mathit{i.e.}$ $\left(C_{\mathrm{m}}(T)=[C_{\mathrm{p}}(T)-C_{\mathrm{lat}}-C_{\mathrm{Sch}}]\right)$,
where $\mathit{C_{\mathrm{lat}}}$ and $\mathit{C_{\mathrm{Sch}}}$
are the lattice and Schottky heat capacities respectively. As there
is no suitable non-magnetic analog, we attempted to fit the specific
heat capacity data with a combination of one Debye term $\left(C_{\mathrm{d}}\left[9nR(\frac{T}{\theta_{D}})^{3}\intop_{0}^{x_{D}}\frac{x^{4}e^{x}}{(e^{x}-1)^{2}}dx\right]\right)$
and several Einstein terms $\left(\sum C_{\mathrm{e}_{i}}\left[3nR(\frac{\theta_{E_{i}}}{T})^{2}\frac{exp(\frac{\theta_{E_{i}}}{T})}{(exp(\frac{\theta_{E_{i}}}{T})-1)^{2}}\right]\right)$to
determine the $\mathit{C_{\mathrm{lat}}}$. Among them, one Debye
function plus three Einstein functions (1D+3E) fit was the best in
the fit range 18 - 135\,K. Here the coefficient $C_{\mathrm{d}}$
is the relative weight of the acoustic modes of vibration and coefficients
$C_{\mathrm{e1}}$, $C_{\mathrm{e2}}$ and $C_{\mathrm{e3}}$ are
the relative weights of the optical modes of vibrations. After fitting
we obtained $C_{\mathrm{d}}$:$C_{\mathrm{e1}}$:$C_{\mathrm{e2}}$:$C_{\mathrm{e3}}$
= 1:6:4:3. The sum of these coefficients is equal to the total number
of atoms (n = 14) per formula unit of LZVO. 
\begin{figure}[h]
\centering{}\includegraphics[scale=0.48]{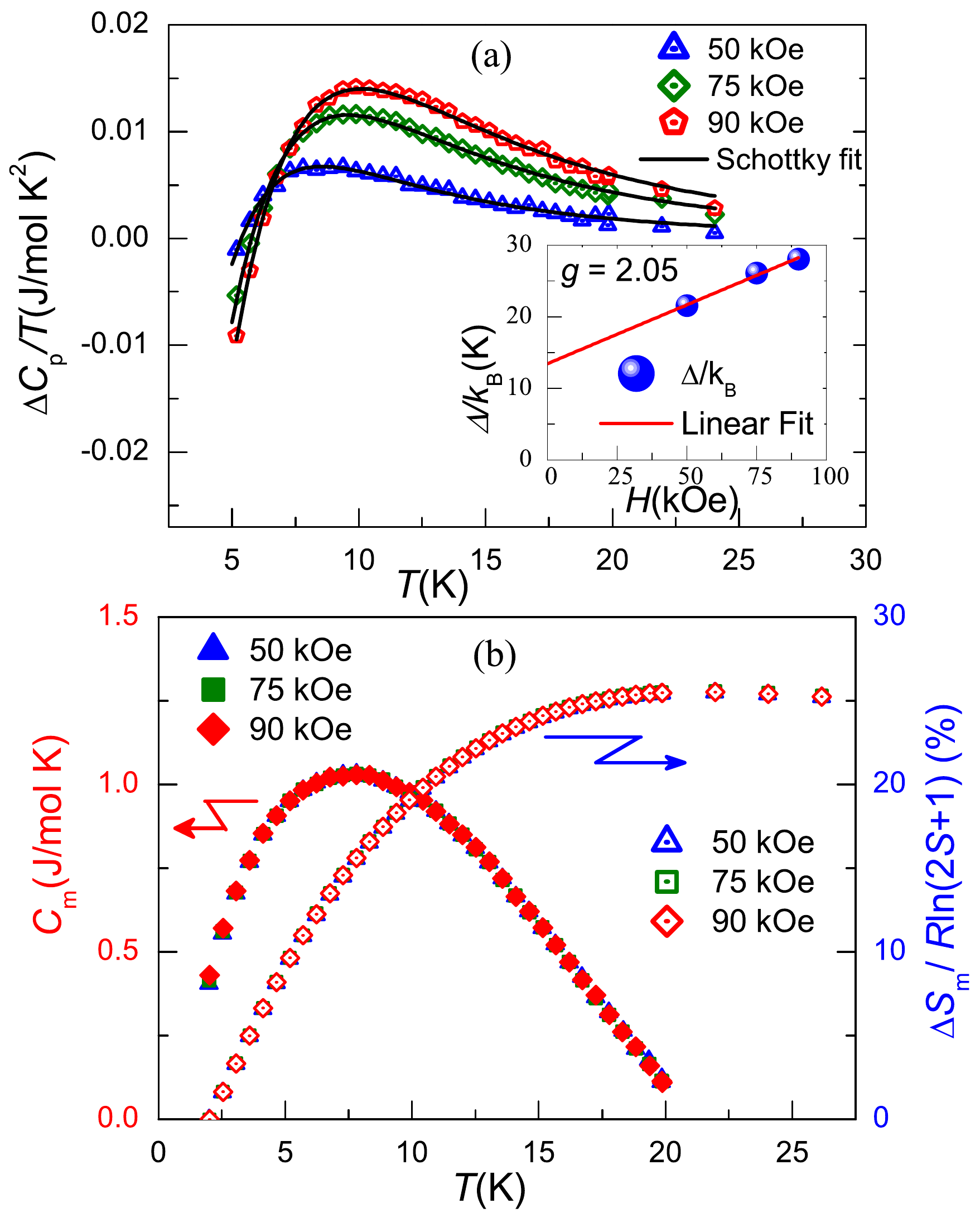}\caption{\label{fig:Csch of LZVO}{\small{}(a) It shows the low temperature
Schottky anomaly fitting in different fields. The inset shows the
linear dependence of the Schottky gap $(\varDelta$) with the field
strength $(\mathit{H}$). (b) The magnetic heat capacity ($C_{\mathrm{m}}$)
on left $\mathit{y}$-axis and the magnetic entropy change ($\Delta S_{\mathrm{m}}$)
on right $\mathit{y}$-axis of LZVO as a function of temperature
is shown.}}
\end{figure}
In the absence of an applied field, the $\mathit{C_{\mathrm{p}}}(T)$
data at $\mathit{H}$ = 0 kOe, $\mathit{C_{\mathrm{p}}\mathrm{(}\mathrm{0},T\mathrm{)}}$
contains lattice $\mathit{C_{\mathrm{lat}}}(T)$, as well as a magnetic
$\mathit{C_{\mathrm{m}}}$$\mathit{\mathrm{(}H,T})$ part but lacks
a Schottky contribution. So we consider $\mathit{C_{\mathrm{p}}\mathrm{(}\mathrm{0},T\mathrm{)}}$
as a reference and subtract it from the higher field data $\mathit{C_{\mathrm{p}}\mathrm{(}H,T}$)
to obtain Schottky contribution in that particular field, $C_{\mathrm{Sch}}(H,T)=$
{[}$C_{\mathrm{p}}(H,T)-C_{\mathrm{p}}(0,T)]=\Delta C_{\mathrm{p}}$.
Now this $\Delta C_{\mathrm{p}}$ represents the Schottky contribution
provided that $\mathit{\mathit{C_{\mathrm{m}}}\mathrm{(}H,T\mathrm{)}}$
is field independent in that temperature range. In Fig. \ref{fig:Csch of LZVO}(a).
we have plotted $\Delta C_{\mathrm{p}}/T$ vs. $\mathit{T}$ and the
Schottky contributions ($C_{\mathrm{Sch}})$ to the total heat capacity are well fitted by a two level Schottky system $C_{Sch}=f\left[R(\frac{\varDelta}{k_{B}T})^{2}(\frac{g_{0}}{g_{1}})\frac{exp(\frac{\varDelta}{k_{B}T})}{[1+(\frac{g_{0}}{g_{1}})exp(\frac{\varDelta}{k_{B}T})]^{2}}\right]$
\citep{Gopal1966,Kumar2016}; where $\mathit{f}$ is the fraction
of free spins within the system, $\varDelta$ is the Schottky gap, $\mathit{R}$
is the universal gas constant, $k_{B}$ is the Boltzmann constant
and $g_{0}$ and $g_{1}$ are the degeneracies of the ground state
and excited state, respectively (in this case, $\mathit{g=g_{0}=\mathrm{1}}$).
Here, with respect to the 25 kOe data, we have derived the Schottky
contribution at higher fields as the zero field data have the same
kind of anomaly present which might be a result of some intrinsic
interactions present within the system. From the fitting of $\Delta C_{\mathrm{p}}(T)/T$,
we obtained the energy gap $\varDelta$ at different fields which are
plotted in the inset of Fig. \ref{fig:Csch of LZVO}(a) and show
a linear dependence. From the slope of $\varDelta/k_{B}$ vs. $\mathit{H}$
plot, we obtained the value of Land$\mathrm{\acute{e}}$ $\mathit{g}$
factor, $\mathit{g}$ = 2.05 which is close to the value for free
spin. From our fits, the fraction of $\mathit{S}$ = $\frac{1}{2}$
entities contributing to the heat capacity is about 2-5 \%. The large
intercept in the $\varDelta/k_{B}$ vs. $\mathit{H}$ shows that the
internal magnetism exists even in zero applied field as might be expected
in a spin-glass. After subtracting the lattice and Schottky contribution,
we obtained the magnetic heat capacity $\mathit{C_{\mathrm{m}}}$
at 50 kOe, 75 kOe and 90 kOe. Fig. \ref{fig:Csch of LZVO}(b) shows
the magnetic heat capacity ($\mathit{C_{\mathrm{m}}}$) on the left
$\mathit{y}$-axis. The $\mathit{C_{\mathrm{m}}}$ is independent
of the applied field and it shows a hump around 7.5 K. Fig. \ref{fig:Csch of LZVO}(b)
also shows the derived magnetic entropy change ($\Delta S_{\mathrm{m}}(T)$)
on the right $\mathit{y}$-axis. The maximum value of $\Delta S_{\mathrm{m}}(T)$
(=$\intop\frac{C_{\mathrm{m}}}{T}dT$) is 1.75 (J/mol K) which is only 25\% of the expected $\mathit{Rln\mathrm{(}2S+\mathrm{1}\mathrm{)}}$
= 6.89 (J/mol K) for two $\mathit{S}$ = $\frac{1}{2}$ spin ($\mathrm{V}^{4+}$
ions) and one $\mathit{S}$ = 1 spin ($\mathrm{V}^{3+}$ ion) per
formula unit of LZVO. A large reduction of entropy change indicates
quenching of moments due to frustration and/or the presence of many
degenerate low-energy states at low temperature. 
\begin{figure}[h]
\centering{}\includegraphics[scale=0.4]{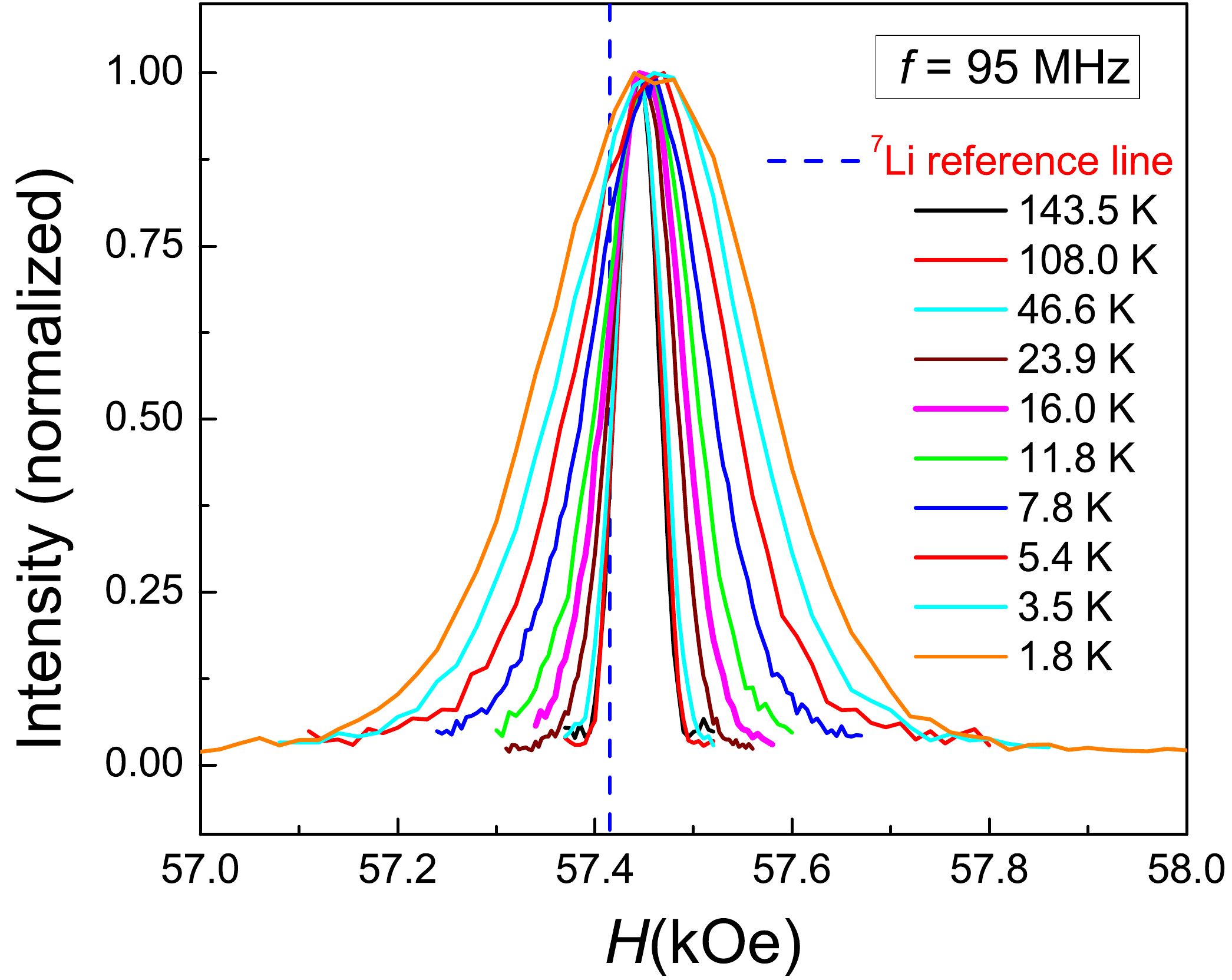}\caption{\label{fig:-NMR-Spectra_LZVO}{\small{}The $^{7}\mathrm{Li}$ NMR
spectra of LZVO at different temperatures measured at a fixed frequency
of 95 MHz. The blue dashed vertical line represents the reference
position for $^{7}$Li nuclei at RT.}}
\end{figure}

\begin{center}
\textbf{\large{}G. $^{7}$Li NMR spectra and shift}{\large\par}
\par\end{center}

In LZVO, the non-magnetic Li ions and magnetic vanadium ions share
the B-site. Due to a strong onsite coupling between the nuclear moment
and the electronic moment it is difficult to detect NMR signal
from the magnetic vanadium ions yielding very short relaxation times.
For Li$^{1+}$ ions, the electronic spin is zero ($\mathit{S}$ =
0), with a nuclear spin $\mathit{I}$ = $\frac{3}{2}$. Additionally,
the $^{7}$Li nuclei exhibit a high natural abundance (92.6\%) and
a high gyromagnetic ratio $\frac{\gamma}{2\pi}$ = 1.6546 MHz/kOe
resulting in strong nuclear magnetization detectable in NMR experiments.
$^{7}$Li NMR spectra have been measured at 95 MHz down to 1.8 K.
The spin-lattice and spin-spin relaxation rates also have been measured
to probe the low energy excitations.

\begin{figure}[h]
\begin{centering}
\includegraphics[scale=0.36]{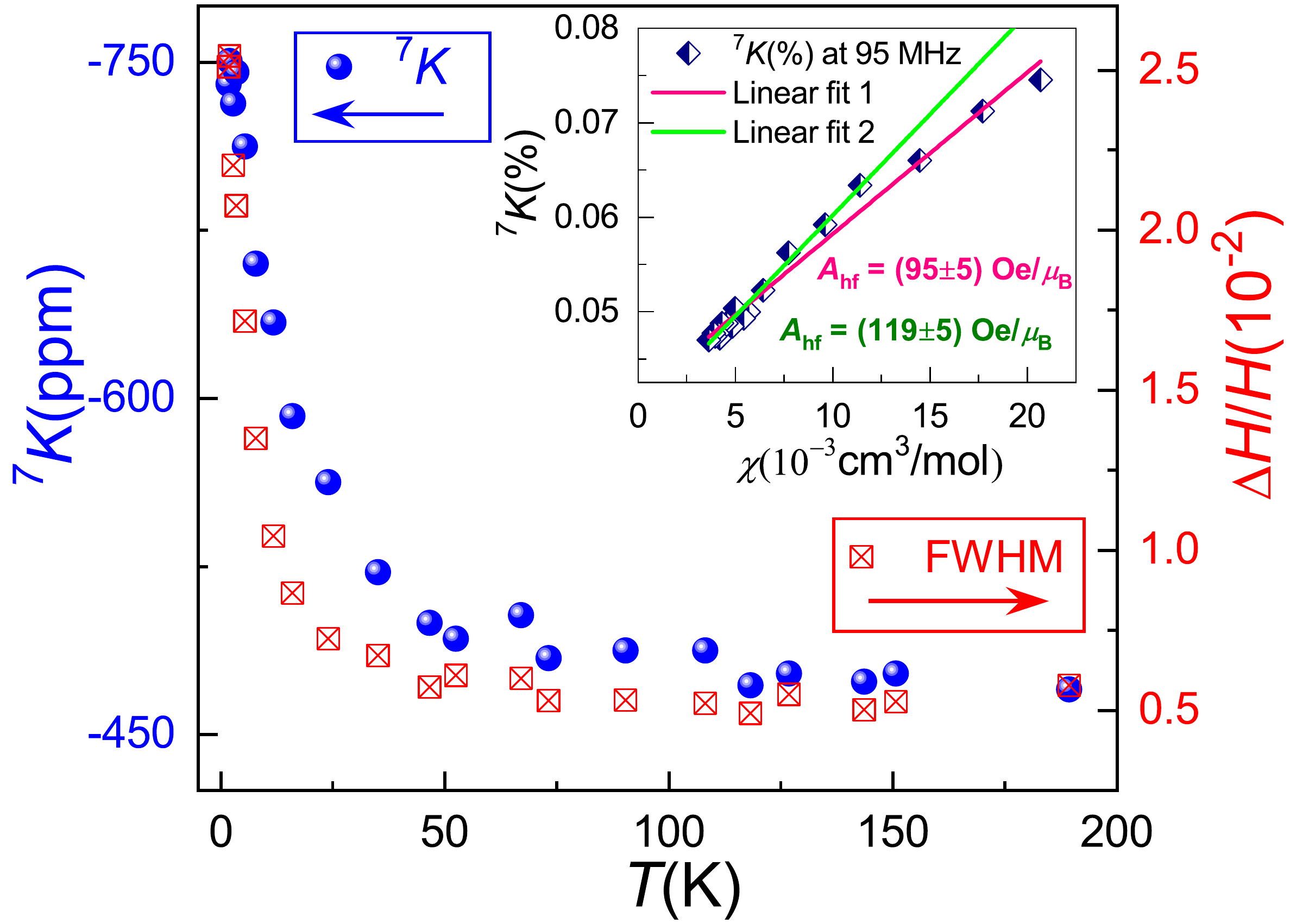}\caption{\label{fig:The-Knight-shift}{\small{}The $^{7}$Li NMR shift (left
$\mathit{y}$-axis) and the FWHM (right $\mathit{y}$-axis) of the
$^{7}$Li NMR spectra as a function of temperature behave like the
paramagnetic bulk susceptibility. In the inset, the linear variation
of }$\mathit{^{7}K\mathrm{(\%)}}$ vs. $\mathit{\chi}${\small{} with
temperature as an implicit parameter is shown and from the slope we
determine the hyperfine coupling constant $\mathit{A_{\mathrm{hf}}}$.}}
\par\end{centering}
\end{figure}

Fig. \ref{fig:-NMR-Spectra_LZVO} shows the spectra at different temperatures
from 143 K to 1.8 K. The spectra get broadened and nominally shift
towards higher fields ($\mathit{i.e.}$ negative shift concerning frequency)
as temperature decreases. By fitting the spectra at different temperatures
to Gaussian functions we derived the full width at half maximum (FWHM)
and NMR line shift $\mathit{^{7}K}$ (using the relation $^{7}K$
= $\frac{H_{ref}-H_{res}}{H_{ref}}$; where $\mathit{H_{res}}$ denotes
the resonance field which is estimated by considering the peak position
of that spectrum, while $\mathit{H_{ref}}$ represents the reference
field of the $^{7}$Li nuclei at room temperature). The calculated
shift ($^{7}K$) varies from -450 ppm (200 K) to -750 ppm (1.8 K)
with respect to the $^{7}$Li NMR reference field $\mathit{H_{ref}}$
= 57.415 kOe at 95 MHZ. The temperature dependence of FWHM and $^{7}K$
are shown in Fig. \ref{fig:The-Knight-shift} on the right and left
$\mathit{y}$-axis, respectively. Both exhibit paramagnetic behavior
according to the Curie-Weiss law of the bulk susceptibility $\chi(T)$.
In the inset of Fig. \ref{fig:The-Knight-shift}, we have plotted
the $^{7}K$(\%) vs. dc susceptibility $\chi(T)$ with temperature
as an implicit parameter and the linear behavior is seen. As $K_{\mathrm{shift}}=\frac{A_{\mathrm{hf}}}{N_{\mathrm{A}}\mu_{\mathrm{B}}}\chi(T)$;
the slope of the linear fit gives the hyperfine coupling constant
($\mathit{A_{\mathrm{hf}}}$). We have fitted the $\mathit{^{7}K}$-$\chi$
plot by a linear equation in two different ranges which are designated
as linear fit 1 and linear fit 2. In linear fit 1, we have considered
the whole temperature range and it gives us $\mathit{A_{\mathrm{hf}}}$
= (95$\pm5$) Oe/$\mathrm{\mu_{B}}$. On the other hand, in linear
fit 2, we have neglected a few low-temperature points (as, here, there
could be some extrinsic Curie contribution due to paramagnetic impurities
in the bulk susceptibility data) and this fit yields $\mathit{A_{\mathrm{hf}}}$
= (119$\pm5$) Oe/$\mathrm{\mu_{B}}.$

\begin{figure}[h]
\centering{}\includegraphics[scale=0.35]{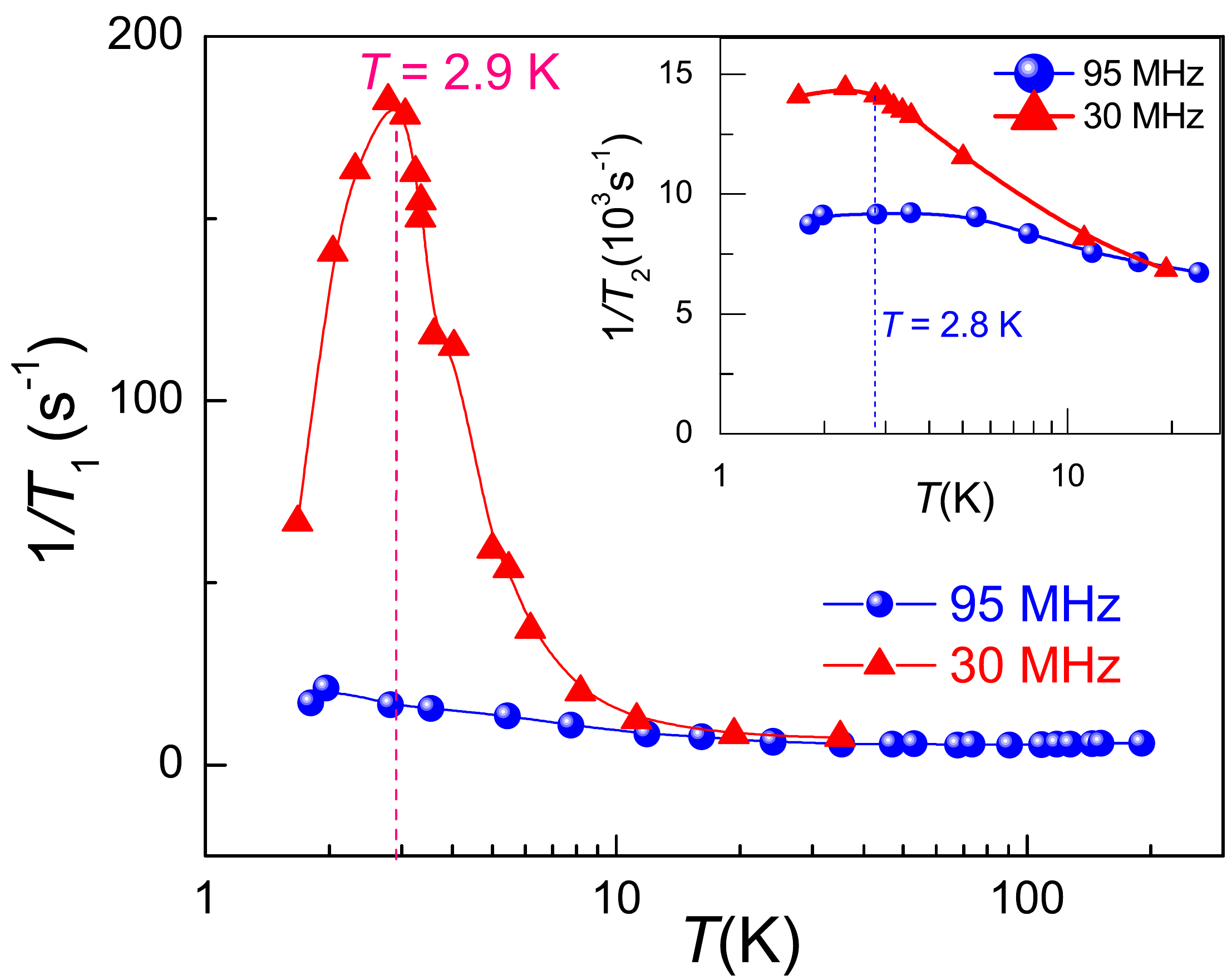}\caption{\label{fig:1/T1 of LZVO}{\small{}The spin-lattice relaxation rate
(}$\frac{1}{T_{1}}${\small{}) and (in the inset) the spin-spin relaxation
rate (}$\frac{1}{T_{2}}${\small{}) of $^{7}$Li are shown in semi-log
scale. Both were measured at two different frequencies (95 MHz and
30 MHz) with a prominent peak at around $T_{f}$ $\simeq$ 3 K in lower frequency
according to a lower applied magnetic field. The connecting lines
are a guides to the eye.}}
\end{figure}

\begin{center}
\textbf{\large{}H. Spin-lattice and spin-spin relaxation} 
\par\end{center}

\begin{figure}[h]
\centering{}\includegraphics[scale=0.4]{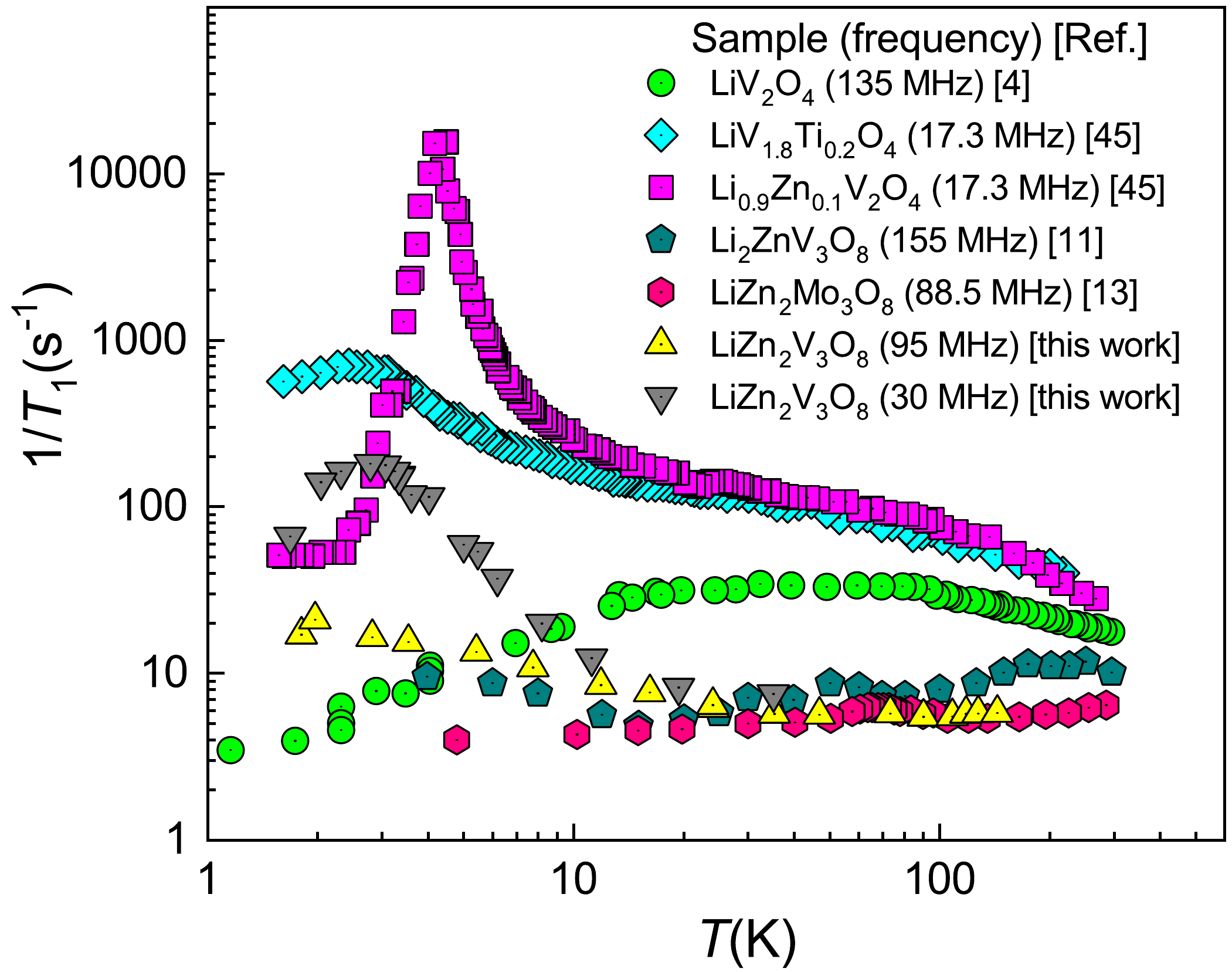}\caption{{\small{}\label{fig:compare1/T1}The spin-lattice relaxation rate
(1/$\mathit{T_{\mathrm{1}}}$) of $^{7}$Li nuclei in LZVO is compared
to the pure and doped LiV$_{2}$O$_{4}$ systems from literature.}}
\end{figure}

The spin-lattice relaxation time ($T_{1}$) of $^{7}$Li at various
temperatures was measured using a saturation recovery sequence at
two fixed frequencies 95 MHz and 30 MHz. The data are well fitted
with a single exponential $\left(1-\frac{M(t)}{M(0)}=Ae{}^{(\frac{-t}{T_{1}})}\right)$
in the temperature range (150 - 7 K). Around 3 K the data are best
fitted with a stretched exponential. At low fields, the spin-lattice
relaxation rate shows a prominent peak around 2.9 K which is close
to the spin-glass ordering temperature $T_{\mathrm{\mathit{f}}}$
= 3 K (see Fig. \ref{fig:1/T1 of LZVO}). This is in agreement with
the bulk dc susceptibility, where, in high fields, the anomaly is suppressed.
The spin-spin relaxation data are well fitted with the single exponential
$\left(M(t)=M_{0}e{}^{(\frac{-2t}{T_{2}})}\right)$. The spin-spin
relaxation rates $(\frac{1}{T_{2}})$ (see inset of Fig. \ref{fig:1/T1 of LZVO})
also show an anomaly around the SG transition temperature $T_{f}$
$\simeq$ 3 K. The $^{7}$Li NMR $\frac{1}{T_{1}}$ data at high-field is nearly
unchanged with temperature, which is different from the published
data of undoped LiV$_{2}$O$_{4}$ \citep{Kondo1997} and doped LiV$_{2}$O$_{4}$
\citep{Brando2002} (see Fig. \ref{fig:compare1/T1} where published
data are shown along with our data). In low-field $\mathit{i.e}$.
at 30 MHz, we observed an increase in our $\frac{1}{T_{1}}$ data
of $^{7}$Li nuclei for LZVO near the freezing temperature $\mathit{T_{f}}$
= 3 K (see Fig. \ref{fig:1/T1 of LZVO}). Likewise in case of Li$_{1-x}$Zn$_{x}$V$_{2}$O$_{4}$
($\mathit{x}$ = 0.1) and Li(V$_{1-y}$Ti$_{y}$)$_{2}$O$_{4}$ ($\mathit{y}$
= 0.1) \citep{Trinkl2000}, an anomaly or peak in the $\mathit{T}$-dependence
of $^{7}$Li NMR $\frac{1}{T_{1}}$ was seen close to the SG transition
temperature. Also, $^{7}$Li NMR studies of $\mathit{S}$ = $\frac{1}{2}$
geometrically frustrated systems, Li$_{2}$ZnV$_{3}$O$_{8}$ \citep{Chakrabarty2014a}
and LiZn$_{2}$Mo$_{3}$O$_{8}$ \citep{Sheckelton2014} have been
reported with an almost temperature invariant spin-lattice relaxation
rate ($\frac{1}{T_{1}}$) of the order of 10 s$^{-1}$ similar to
what we have seen for our LZVO at 95 MHz. As the earlier reported
spin-lattice relaxation data on pure and doped LiV$_{2}$O$_{4}$
systems are measured at different frequencies, we have mentioned them
in bracket for clarity in Fig. \ref{fig:compare1/T1}. 

\section{Discussion}

\begin{figure}[h]
\centering{}\includegraphics[scale=0.28]{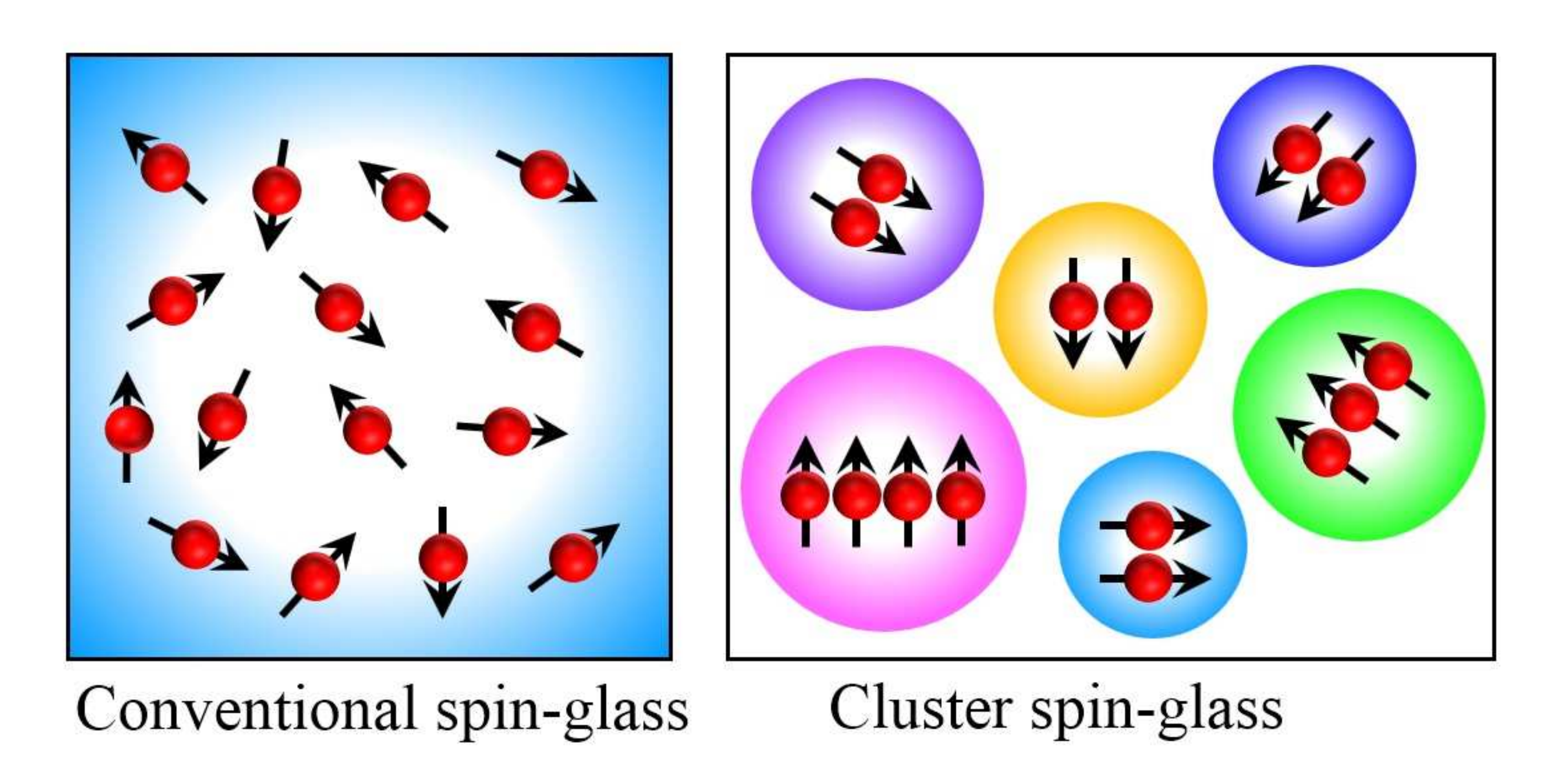}\caption{\label{fig:Illustration-of-CSG-1} Illustration of conventional spin-glass
and cluster spin-glass. The colors indicate the different domains
formed by group of spins which are oriented in a particular direction
with different magnetic strength.}
\end{figure}

Theoretical studies by M. Schmidt \textit{et al}. \citep{Schmidt2017},
Gang Chen \textit{et al}. \citep{Chen2014,Chen2016,Chen2018} and
E. C. Andrade \textit{et al}. \citep{Andrade2018} show that when
a geometrically frustrated system (like the Kagom$\mathrm{\acute{e}}$
lattice) is driven by disorder then it often manifests spin-glass
(SG) and sometimes cluster-spin-glass (CSG) properties. A sketch of
the outcome is given in Fig. \ref{fig:Illustration-of-CSG-1} which
depicts a cartoon of spin orientation in a conventional
SG and a CSG. The individual spins are locked in place for conventional
SG systems whereas in CSG, a group of spins are locked to form a pair,
triplet or even domains. The remaining spins which are not participating
in the cluster are independent, but help to mediate the interactions
between the clusters such that the clusters can change their sizes
and response time. The bigger the cluster size, the slower is the
relaxation rate. The clusters can be short-range ferromagnetic in nature.
But these need not to be fully FM entities; they can be mostly random with
some correlation length which rigidly couples more than two spins.

Among the spinels where the pyrochlore lattice of the magnetic B-sites
is diluted by non-magnetic ions in the ratio 1:3, for LiZn$_{2}$V$_{3}$O$_{8}$
the Curie-Weiss temperature ($\theta_{\mathrm{CW}}$ = -185 K) is
somewhat lower than in Li$_{2}$ZnV$_{3}$O$_{8}$ $(\theta_{\mathrm{CW}}$
= -214 K) and much lower compared to Zn$_{3}$V$_{3}$O$_{8}$ ($\theta_{\mathrm{CW}}$
= -370 K). In all cases, the disorder at B-site of the spinel and
the dilution of the corner-shared tetrahedral network in 3D due to
the presence of 25\% non-magnetic ions (Zn$^{2+}$ for Zn$_{3}$V$_{3}$O$_{8}$,
Li$^{1+}$ for Li$_{2}$ZnV$_{3}$O$_{8}$ and LiZn$_{2}$V$_{3}$O$_{8}$)
triggers the system to form a spin\textendash frozen state (cluster
spin-glass rather than a conventional one) at about the same temperature.
Considering the ratio of V$^{3+}$ and V$^{4+}$ ions, the expected
Curie constant per vanadium is $\mathit{C}$ = 0.79 Kcm$^{3}$/mol
for ZVO and $\mathit{C}$ = 0.58 Kcm$^{3}$/mol for LZVO; whereas
the obtained values are $\mathit{C}$ = 0.75 Kcm$^{3}$/mol and $\mathit{C}$
= 0.28 Kcm$^{3}$/mol, respectively. We speculate that the much lower
Curie constant in LZVO might indicate the presence of some degree
of itinerancy like in the metallic LiV$_{2}$O$_{4}$ system. It would
be interesting to see how these parameters evolve with the extent
of dilution.

\section{Conclusion}

The cubic spinel LiZn$_{2}$V$_{3}$O$_{8}$ has been successfully
synthesized. It crystallizes in the centrosymmetric cubic spinel $\mathit{Fd\bar{\mathrm{3}}m}$
space group. From a CW-fit of our 10 kOe dc susceptibility data, we
obtained the Curie constant $\mathit{C}$ = 0.28 Kcm$^{3}$/mol V
and CW temperature $\theta_{\mathrm{CW}}$ = -185 K. The Curie constant
suggests a possible suppression or partial quenching of local moments
and the CW temperature suggests strong AFM interactions among the
magnetic vanadium ions. A high value of frustration parameter ($\mathit{f}$
$\simeq $ 60) is inferred from our data. The ZFC-FC bifurcation below $\mathit{T_{f}}\simeq$ 3 K indicates SG nature which is confirmed by frequency dependence
of the ac susceptibility, the Vogel-Fulcher law and critical power
law of the $\mathit{T_{f}}$. The low value of the characteristic
angular frequency $\omega_{0}$ $\approx$ 3.56$\times$10$^{6}$
Hz and the high value of the critical time constant $\tau_{0}$ $\approx$
1.82$\times$10$^{-6}$ s, corroborates the CSG ground state. The
magnetic relaxation, aging phenomena and memory effect support the
metastability of the CSG ground state. The asymmetric response in
a positive temperature cycle obeys the hierarchical model proposed
for the SG systems. Absence of any sharp anomaly in heat capacity
data indicates lack of long-range ordering down to 2 K. The significant
contribution of magnetic heat capacity shows a hump around 7.5 K which
is independent of the applied field strength. The small entropy change
$(\Delta S_{\mathrm{m}}$ $\simeq25\%$) presumably arises from a  large degeneracy of the ground states. The
field swept $^{7}$Li NMR spectra show a line shift ($^{7}K$) as
$\mathit{T}$ decreases. The plot of line shift $^{7}K$(\%) vs. bulk
susceptibility $\chi$ follows a linear behavior and the derived hyperfine
coupling constant amounts to $\mathit{A_{\mathrm{hf}}}$ = 119 Oe/$\mu_{\mathrm{B}}$.
The temperature dependence of the spin-lattice relaxation rate ($\frac{1}{T_{1}}$),
and the spin-spin relaxation rate ($\frac{1}{T_{2}}$) of the $^{7}$Li
nuclei indicates that they are sensitive to the fluctuations of the
magnetic ions and show an anomaly near $T_{f}$. All experimental
results and observations point towards the formation of a CSG ground
state in LZVO.

\section{Acknowledgments}

SK acknowledges the central facility and financial support from IRCC,
IIT Bombay. AVM would like to thank the Alexander von Humboldt foundation
for financial support during his stay at Augsburg Germany. We kindly
acknowledge support from the Deutsche Forschungsgemeinschaft (DFG,
German Research Foundation) \textendash{} Projektnummer 107745057
\textendash{} TRR 80.

\bibliographystyle{apsrev4-1}
\bibliography{Citation}

\begin{thebibliography}{50}%
\makeatletter
\providecommand \@ifxundefined [1]{%
 \@ifx{#1\undefined}
}%
\providecommand \@ifnum [1]{%
 \ifnum #1\expandafter \@firstoftwo
 \else \expandafter \@secondoftwo
 \fi
}%
\providecommand \@ifx [1]{%
 \ifx #1\expandafter \@firstoftwo
 \else \expandafter \@secondoftwo
 \fi
}%
\providecommand \natexlab [1]{#1}%
\providecommand \enquote  [1]{``#1''}%
\providecommand \bibnamefont  [1]{#1}%
\providecommand \bibfnamefont [1]{#1}%
\providecommand \citenamefont [1]{#1}%
\providecommand \href@noop [0]{\@secondoftwo}%
\providecommand \href [0]{\begingroup \@sanitize@url \@href}%
\providecommand \@href[1]{\@@startlink{#1}\@@href}%
\providecommand \@@href[1]{\endgroup#1\@@endlink}%
\providecommand \@sanitize@url [0]{\catcode `\\12\catcode `\$12\catcode
  `\&12\catcode `\#12\catcode `\^12\catcode `\_12\catcode `\%12\relax}%
\providecommand \@@startlink[1]{}%
\providecommand \@@endlink[0]{}%
\providecommand \url  [0]{\begingroup\@sanitize@url \@url }%
\providecommand \@url [1]{\endgroup\@href {#1}{\urlprefix }}%
\providecommand \urlprefix  [0]{URL }%
\providecommand \Eprint [0]{\href }%
\providecommand \doibase [0]{http://dx.doi.org/}%
\providecommand \selectlanguage [0]{\@gobble}%
\providecommand \bibinfo  [0]{\@secondoftwo}%
\providecommand \bibfield  [0]{\@secondoftwo}%
\providecommand \translation [1]{[#1]}%
\providecommand \BibitemOpen [0]{}%
\providecommand \bibitemStop [0]{}%
\providecommand \bibitemNoStop [0]{.\EOS\space}%
\providecommand \EOS [0]{\spacefactor3000\relax}%
\providecommand \BibitemShut  [1]{\csname bibitem#1\endcsname}%
\let\auto@bib@innerbib\@empty
\bibitem [{\citenamefont {Greedan}(2001)}]{Greedan2001}%
  \BibitemOpen
  \bibfield  {author} {\bibinfo {author} {\bibfnamefont {J.~E.}\ \bibnamefont
  {Greedan}},\ }\href {\doibase 10.1039/B003682J} {\bibfield  {journal}
  {\bibinfo  {journal} {J. Mater. Chem.}\ }\textbf {\bibinfo {volume} {11}},\
  \bibinfo {pages} {37} (\bibinfo {year} {2001})}\BibitemShut {NoStop}%
\bibitem [{\citenamefont {Moessner}\ and\ \citenamefont
  {Ramirez}(2006)}]{Moessner2006}%
  \BibitemOpen
  \bibfield  {author} {\bibinfo {author} {\bibfnamefont {R.}~\bibnamefont
  {Moessner}}\ and\ \bibinfo {author} {\bibfnamefont {A.~P.}\ \bibnamefont
  {Ramirez}},\ }\href {\doibase 10.1063/1.2186278} {\bibfield  {journal}
  {\bibinfo  {journal} {Physics Today}\ }\textbf {\bibinfo {volume} {59}},\
  \bibinfo {pages} {24} (\bibinfo {year} {2006})}\BibitemShut {NoStop}%
\bibitem [{\citenamefont {Chmaissem}\ \emph {et~al.}(1997)\citenamefont
  {Chmaissem}, \citenamefont {Jorgensen}, \citenamefont {Kondo},\ and\
  \citenamefont {Johnston}}]{Chmaissem1997}%
  \BibitemOpen
  \bibfield  {author} {\bibinfo {author} {\bibfnamefont {O.}~\bibnamefont
  {Chmaissem}}, \bibinfo {author} {\bibfnamefont {J.~D.}\ \bibnamefont
  {Jorgensen}}, \bibinfo {author} {\bibfnamefont {S.}~\bibnamefont {Kondo}}, \
  and\ \bibinfo {author} {\bibfnamefont {D.~C.}\ \bibnamefont {Johnston}},\
  }\href {\doibase 10.1103/PhysRevLett.79.4866} {\bibfield  {journal} {\bibinfo
   {journal} {Phys. Rev. Lett.}\ }\textbf {\bibinfo {volume} {79}},\ \bibinfo
  {pages} {4866} (\bibinfo {year} {1997})}\BibitemShut {NoStop}%
\bibitem [{\citenamefont {Kondo}\ \emph {et~al.}(1997)\citenamefont {Kondo},
  \citenamefont {Johnston}, \citenamefont {Swenson}, \citenamefont {Borsa},
  \citenamefont {Mahajan}, \citenamefont {Miller}, \citenamefont {Gu},
  \citenamefont {Goldman}, \citenamefont {Maple}, \citenamefont {Gajewski},
  \citenamefont {Freeman}, \citenamefont {Dilley}, \citenamefont {Dickey},
  \citenamefont {Merrin}, \citenamefont {Kojima}, \citenamefont {Luke},
  \citenamefont {Uemura}, \citenamefont {Chmaissem},\ and\ \citenamefont
  {Jorgensen}}]{Kondo1997}%
  \BibitemOpen
  \bibfield  {author} {\bibinfo {author} {\bibfnamefont {S.}~\bibnamefont
  {Kondo}}, \bibinfo {author} {\bibfnamefont {D.~C.}\ \bibnamefont {Johnston}},
  \bibinfo {author} {\bibfnamefont {C.~A.}\ \bibnamefont {Swenson}}, \bibinfo
  {author} {\bibfnamefont {F.}~\bibnamefont {Borsa}}, \bibinfo {author}
  {\bibfnamefont {A.~V.}\ \bibnamefont {Mahajan}}, \bibinfo {author}
  {\bibfnamefont {L.~L.}\ \bibnamefont {Miller}}, \bibinfo {author}
  {\bibfnamefont {T.}~\bibnamefont {Gu}}, \bibinfo {author} {\bibfnamefont
  {A.~I.}\ \bibnamefont {Goldman}}, \bibinfo {author} {\bibfnamefont {M.~B.}\
  \bibnamefont {Maple}}, \bibinfo {author} {\bibfnamefont {D.~A.}\ \bibnamefont
  {Gajewski}}, \bibinfo {author} {\bibfnamefont {E.~J.}\ \bibnamefont
  {Freeman}}, \bibinfo {author} {\bibfnamefont {N.~R.}\ \bibnamefont {Dilley}},
  \bibinfo {author} {\bibfnamefont {R.~P.}\ \bibnamefont {Dickey}}, \bibinfo
  {author} {\bibfnamefont {J.}~\bibnamefont {Merrin}}, \bibinfo {author}
  {\bibfnamefont {K.}~\bibnamefont {Kojima}}, \bibinfo {author} {\bibfnamefont
  {G.~M.}\ \bibnamefont {Luke}}, \bibinfo {author} {\bibfnamefont {Y.~J.}\
  \bibnamefont {Uemura}}, \bibinfo {author} {\bibfnamefont {O.}~\bibnamefont
  {Chmaissem}}, \ and\ \bibinfo {author} {\bibfnamefont {J.~D.}\ \bibnamefont
  {Jorgensen}},\ }\href {\doibase 10.1103/PhysRevLett.78.3729} {\bibfield
  {journal} {\bibinfo  {journal} {Phys. Rev. Lett.}\ }\textbf {\bibinfo
  {volume} {78}},\ \bibinfo {pages} {3729} (\bibinfo {year}
  {1997})}\BibitemShut {NoStop}%
\bibitem [{\citenamefont {Krimmel}\ \emph {et~al.}(2004)\citenamefont
  {Krimmel}, \citenamefont {Loidl}, \citenamefont {Klemm}, \citenamefont
  {Horn}, \citenamefont {Sheptyakov},\ and\ \citenamefont
  {Fischer}}]{Krimmel2004}%
  \BibitemOpen
  \bibfield  {author} {\bibinfo {author} {\bibfnamefont {A.}~\bibnamefont
  {Krimmel}}, \bibinfo {author} {\bibfnamefont {A.}~\bibnamefont {Loidl}},
  \bibinfo {author} {\bibfnamefont {M.}~\bibnamefont {Klemm}}, \bibinfo
  {author} {\bibfnamefont {S.}~\bibnamefont {Horn}}, \bibinfo {author}
  {\bibfnamefont {D.~V.}\ \bibnamefont {Sheptyakov}}, \ and\ \bibinfo {author}
  {\bibfnamefont {P.}~\bibnamefont {Fischer}},\ }\href {\doibase
  https://doi.org/10.1016/j.physb.2004.03.074} {\bibfield  {journal} {\bibinfo
  {journal} {Physica B: Condensed Matter}\ }\textbf {\bibinfo {volume} {350}},\
  \bibinfo {pages} {E297 } (\bibinfo {year} {2004})},\ \bibinfo {note}
  {proceedings of the Third European Conference on Neutron
  Scattering}\BibitemShut {NoStop}%
\bibitem [{\citenamefont {Stewart}(1984)}]{Stewart1984}%
  \BibitemOpen
  \bibfield  {author} {\bibinfo {author} {\bibfnamefont {G.~R.}\ \bibnamefont
  {Stewart}},\ }\href {\doibase 10.1103/RevModPhys.56.755} {\bibfield
  {journal} {\bibinfo  {journal} {Rev. Mod. Phys.}\ }\textbf {\bibinfo {volume}
  {56}},\ \bibinfo {pages} {755} (\bibinfo {year} {1984})}\BibitemShut
  {NoStop}%
\bibitem [{\citenamefont {Trinkl}\ \emph
  {et~al.}(2000{\natexlab{a}})\citenamefont {Trinkl}, \citenamefont {Loidl},
  \citenamefont {Klemm},\ and\ \citenamefont {Horn}}]{Trinkl2000a}%
  \BibitemOpen
  \bibfield  {author} {\bibinfo {author} {\bibfnamefont {W.}~\bibnamefont
  {Trinkl}}, \bibinfo {author} {\bibfnamefont {A.}~\bibnamefont {Loidl}},
  \bibinfo {author} {\bibfnamefont {M.}~\bibnamefont {Klemm}}, \ and\ \bibinfo
  {author} {\bibfnamefont {S.}~\bibnamefont {Horn}},\ }\href {\doibase
  10.1103/PhysRevB.62.8915} {\bibfield  {journal} {\bibinfo  {journal} {Phys.
  Rev. B}\ }\textbf {\bibinfo {volume} {62}},\ \bibinfo {pages} {8915}
  (\bibinfo {year} {2000}{\natexlab{a}})}\BibitemShut {NoStop}%
\bibitem [{\citenamefont {Miyoshi}\ \emph {et~al.}(2002)\citenamefont
  {Miyoshi}, \citenamefont {Ihara}, \citenamefont {Fujiwara},\ and\
  \citenamefont {Takeuchi}}]{Miyoshi2002}%
  \BibitemOpen
  \bibfield  {author} {\bibinfo {author} {\bibfnamefont {K.}~\bibnamefont
  {Miyoshi}}, \bibinfo {author} {\bibfnamefont {M.}~\bibnamefont {Ihara}},
  \bibinfo {author} {\bibfnamefont {K.}~\bibnamefont {Fujiwara}}, \ and\
  \bibinfo {author} {\bibfnamefont {J.}~\bibnamefont {Takeuchi}},\ }\href
  {\doibase 10.1103/PhysRevB.65.092414} {\bibfield  {journal} {\bibinfo
  {journal} {Phys. Rev. B}\ }\textbf {\bibinfo {volume} {65}},\ \bibinfo
  {pages} {092414} (\bibinfo {year} {2002})}\BibitemShut {NoStop}%
\bibitem [{\citenamefont {Ueda}\ \emph {et~al.}(1997)\citenamefont {Ueda},
  \citenamefont {Fujiwara},\ and\ \citenamefont {Yasuoka}}]{Ueda1997}%
  \BibitemOpen
  \bibfield  {author} {\bibinfo {author} {\bibfnamefont {Y.}~\bibnamefont
  {Ueda}}, \bibinfo {author} {\bibfnamefont {N.}~\bibnamefont {Fujiwara}}, \
  and\ \bibinfo {author} {\bibfnamefont {H.}~\bibnamefont {Yasuoka}},\ }\href
  {\doibase 10.1143/JPSJ.66.778} {\bibfield  {journal} {\bibinfo  {journal}
  {Journal of the Physical Society of Japan}\ }\textbf {\bibinfo {volume}
  {66}},\ \bibinfo {pages} {778} (\bibinfo {year} {1997})}\BibitemShut
  {NoStop}%
\bibitem [{\citenamefont {Chakrabarty}\ \emph
  {et~al.}(2014{\natexlab{a}})\citenamefont {Chakrabarty}, \citenamefont
  {Mahajan},\ and\ \citenamefont {Kundu}}]{Chakrabarty2014}%
  \BibitemOpen
  \bibfield  {author} {\bibinfo {author} {\bibfnamefont {T.}~\bibnamefont
  {Chakrabarty}}, \bibinfo {author} {\bibfnamefont {A.~V.}\ \bibnamefont
  {Mahajan}}, \ and\ \bibinfo {author} {\bibfnamefont {S.}~\bibnamefont
  {Kundu}},\ }\href {http://stacks.iop.org/0953-8984/26/i=40/a=405601}
  {\bibfield  {journal} {\bibinfo  {journal} {Journal of Physics: Condensed
  Matter}\ }\textbf {\bibinfo {volume} {26}},\ \bibinfo {pages} {405601}
  (\bibinfo {year} {2014}{\natexlab{a}})}\BibitemShut {NoStop}%
\bibitem [{\citenamefont {Chakrabarty}\ \emph
  {et~al.}(2014{\natexlab{b}})\citenamefont {Chakrabarty}, \citenamefont
  {Mahajan},\ and\ \citenamefont {Koteswararao}}]{Chakrabarty2014a}%
  \BibitemOpen
  \bibfield  {author} {\bibinfo {author} {\bibfnamefont {T.}~\bibnamefont
  {Chakrabarty}}, \bibinfo {author} {\bibfnamefont {A.~V.}\ \bibnamefont
  {Mahajan}}, \ and\ \bibinfo {author} {\bibfnamefont {B.}~\bibnamefont
  {Koteswararao}},\ }\href {http://stacks.iop.org/0953-8984/26/i=26/a=265601}
  {\bibfield  {journal} {\bibinfo  {journal} {Journal of Physics: Condensed
  Matter}\ }\textbf {\bibinfo {volume} {26}},\ \bibinfo {pages} {265601}
  (\bibinfo {year} {2014}{\natexlab{b}})}\BibitemShut {NoStop}%
\bibitem [{\citenamefont {Sheckelton}\ \emph {et~al.}(2012)\citenamefont
  {Sheckelton}, \citenamefont {Neilson}, \citenamefont {Soltan},\ and\
  \citenamefont {McQueen}}]{Sheckelton2012}%
  \BibitemOpen
  \bibfield  {author} {\bibinfo {author} {\bibfnamefont {J.~P.}\ \bibnamefont
  {Sheckelton}}, \bibinfo {author} {\bibfnamefont {J.~R.}\ \bibnamefont
  {Neilson}}, \bibinfo {author} {\bibfnamefont {D.~G.}\ \bibnamefont {Soltan}},
  \ and\ \bibinfo {author} {\bibfnamefont {T.~M.}\ \bibnamefont {McQueen}},\
  }\href {\doibase 10.1038/nmat3329} {\bibfield  {journal} {\bibinfo  {journal}
  {Nature Materials}\ }\textbf {\bibinfo {volume} {11}},\ \bibinfo {pages}
  {493} (\bibinfo {year} {2012})}\BibitemShut {NoStop}%
\bibitem [{\citenamefont {Sheckelton}\ \emph {et~al.}(2014)\citenamefont
  {Sheckelton}, \citenamefont {Foronda}, \citenamefont {Pan}, \citenamefont
  {Moir}, \citenamefont {McDonald}, \citenamefont {Lancaster}, \citenamefont
  {Baker}, \citenamefont {Armitage}, \citenamefont {Imai}, \citenamefont
  {Blundell},\ and\ \citenamefont {McQueen}}]{Sheckelton2014}%
  \BibitemOpen
  \bibfield  {author} {\bibinfo {author} {\bibfnamefont {J.~P.}\ \bibnamefont
  {Sheckelton}}, \bibinfo {author} {\bibfnamefont {F.~R.}\ \bibnamefont
  {Foronda}}, \bibinfo {author} {\bibfnamefont {L.}~\bibnamefont {Pan}},
  \bibinfo {author} {\bibfnamefont {C.}~\bibnamefont {Moir}}, \bibinfo {author}
  {\bibfnamefont {R.~D.}\ \bibnamefont {McDonald}}, \bibinfo {author}
  {\bibfnamefont {T.}~\bibnamefont {Lancaster}}, \bibinfo {author}
  {\bibfnamefont {P.~J.}\ \bibnamefont {Baker}}, \bibinfo {author}
  {\bibfnamefont {N.~P.}\ \bibnamefont {Armitage}}, \bibinfo {author}
  {\bibfnamefont {T.}~\bibnamefont {Imai}}, \bibinfo {author} {\bibfnamefont
  {S.~J.}\ \bibnamefont {Blundell}}, \ and\ \bibinfo {author} {\bibfnamefont
  {T.~M.}\ \bibnamefont {McQueen}},\ }\href {\doibase
  10.1103/PhysRevB.89.064407} {\bibfield  {journal} {\bibinfo  {journal} {Phys.
  Rev. B}\ }\textbf {\bibinfo {volume} {89}},\ \bibinfo {pages} {064407}
  (\bibinfo {year} {2014})}\BibitemShut {NoStop}%
\bibitem [{\citenamefont {Anderson}(1973)}]{Anderson1973}%
  \BibitemOpen
  \bibfield  {author} {\bibinfo {author} {\bibfnamefont {P.}~\bibnamefont
  {Anderson}},\ }\href {\doibase https://doi.org/10.1016/0025-5408(73)90167-0}
  {\bibfield  {journal} {\bibinfo  {journal} {Materials Research Bulletin}\
  }\textbf {\bibinfo {volume} {8}},\ \bibinfo {pages} {153 } (\bibinfo {year}
  {1973})}\BibitemShut {NoStop}%
\bibitem [{\citenamefont {Reuter}\ and\ \citenamefont
  {Colsmann}(1972)}]{Reuter1972}%
  \BibitemOpen
  \bibfield  {author} {\bibinfo {author} {\bibfnamefont {B.}~\bibnamefont
  {Reuter}}\ and\ \bibinfo {author} {\bibfnamefont {G.}~\bibnamefont
  {Colsmann}},\ }\href@noop {} {\bibfield  {journal} {\bibinfo  {journal} {Z.
  anorg. allg. Chem.}\ }\textbf {\bibinfo {volume} {894}},\ \bibinfo {pages}
  {138} (\bibinfo {year} {1972})}\BibitemShut {NoStop}%
\bibitem [{\citenamefont {Lee}\ \emph {et~al.}(2000)\citenamefont {Lee},
  \citenamefont {Broholm}, \citenamefont {Kim}, \citenamefont {Ratcliff},\ and\
  \citenamefont {Cheong}}]{Lee2000}%
  \BibitemOpen
  \bibfield  {author} {\bibinfo {author} {\bibfnamefont {S.-H.}\ \bibnamefont
  {Lee}}, \bibinfo {author} {\bibfnamefont {C.}~\bibnamefont {Broholm}},
  \bibinfo {author} {\bibfnamefont {T.~H.}\ \bibnamefont {Kim}}, \bibinfo
  {author} {\bibfnamefont {W.}~\bibnamefont {Ratcliff}}, \ and\ \bibinfo
  {author} {\bibfnamefont {S.-W.}\ \bibnamefont {Cheong}},\ }\href {\doibase
  10.1103/PhysRevLett.84.3718} {\bibfield  {journal} {\bibinfo  {journal}
  {Phys. Rev. Lett.}\ }\textbf {\bibinfo {volume} {84}},\ \bibinfo {pages}
  {3718} (\bibinfo {year} {2000})}\BibitemShut {NoStop}%
\bibitem [{\citenamefont {Lee}\ \emph {et~al.}(2002)\citenamefont {Lee},
  \citenamefont {Broholm}, \citenamefont {Ratcliff}, \citenamefont
  {Gasparovic}, \citenamefont {Huang}, \citenamefont {Kim},\ and\ \citenamefont
  {Cheong}}]{Lee2002}%
  \BibitemOpen
  \bibfield  {author} {\bibinfo {author} {\bibfnamefont {S.}~\bibnamefont
  {Lee}}, \bibinfo {author} {\bibfnamefont {C.}~\bibnamefont {Broholm}},
  \bibinfo {author} {\bibfnamefont {W.}~\bibnamefont {Ratcliff}}, \bibinfo
  {author} {\bibfnamefont {G.}~\bibnamefont {Gasparovic}}, \bibinfo {author}
  {\bibfnamefont {Q.}~\bibnamefont {Huang}}, \bibinfo {author} {\bibfnamefont
  {T.~H.}\ \bibnamefont {Kim}}, \ and\ \bibinfo {author} {\bibfnamefont
  {S.~W.}\ \bibnamefont {Cheong}},\ }\href {\doibase 10.1038/nature00964}
  {\bibfield  {journal} {\bibinfo  {journal} {Nature}\ }\textbf {\bibinfo
  {volume} {418}},\ \bibinfo {pages} {856} (\bibinfo {year}
  {2002})}\BibitemShut {NoStop}%
\bibitem [{\citenamefont {Lee}\ \emph {et~al.}(2004)\citenamefont {Lee},
  \citenamefont {Louca}, \citenamefont {Ueda}, \citenamefont {Park},
  \citenamefont {Sato}, \citenamefont {Isobe}, \citenamefont {Ueda},
  \citenamefont {Rosenkranz}, \citenamefont {Zschack}, \citenamefont
  {\'I\~niguez}, \citenamefont {Qiu},\ and\ \citenamefont {Osborn}}]{Lee2004}%
  \BibitemOpen
  \bibfield  {author} {\bibinfo {author} {\bibfnamefont {S.-H.}\ \bibnamefont
  {Lee}}, \bibinfo {author} {\bibfnamefont {D.}~\bibnamefont {Louca}}, \bibinfo
  {author} {\bibfnamefont {H.}~\bibnamefont {Ueda}}, \bibinfo {author}
  {\bibfnamefont {S.}~\bibnamefont {Park}}, \bibinfo {author} {\bibfnamefont
  {T.~J.}\ \bibnamefont {Sato}}, \bibinfo {author} {\bibfnamefont
  {M.}~\bibnamefont {Isobe}}, \bibinfo {author} {\bibfnamefont
  {Y.}~\bibnamefont {Ueda}}, \bibinfo {author} {\bibfnamefont {S.}~\bibnamefont
  {Rosenkranz}}, \bibinfo {author} {\bibfnamefont {P.}~\bibnamefont {Zschack}},
  \bibinfo {author} {\bibfnamefont {J.}~\bibnamefont {\'I\~niguez}}, \bibinfo
  {author} {\bibfnamefont {Y.}~\bibnamefont {Qiu}}, \ and\ \bibinfo {author}
  {\bibfnamefont {R.}~\bibnamefont {Osborn}},\ }\href {\doibase
  10.1103/PhysRevLett.93.156407} {\bibfield  {journal} {\bibinfo  {journal}
  {Phys. Rev. Lett.}\ }\textbf {\bibinfo {volume} {93}},\ \bibinfo {pages}
  {156407} (\bibinfo {year} {2004})}\BibitemShut {NoStop}%
\bibitem [{\citenamefont {Reehuis}\ \emph {et~al.}(2003)\citenamefont
  {Reehuis}, \citenamefont {Krimmel}, \citenamefont {B\"uttgen}, \citenamefont
  {Loidl},\ and\ \citenamefont {Prokofiev}}]{Reehuis2003}%
  \BibitemOpen
  \bibfield  {author} {\bibinfo {author} {\bibfnamefont {M.}~\bibnamefont
  {Reehuis}}, \bibinfo {author} {\bibfnamefont {A.}~\bibnamefont {Krimmel}},
  \bibinfo {author} {\bibfnamefont {N.}~\bibnamefont {B\"uttgen}}, \bibinfo
  {author} {\bibfnamefont {N.~A.}\ \bibnamefont {Loidl}}, \ and\ \bibinfo
  {author} {\bibfnamefont {A.}~\bibnamefont {Prokofiev}},\ }\href {\doibase
  10.1140/epjb/e2003-00282-4} {\bibfield  {journal} {\bibinfo  {journal} {Eur.
  Phys. J. B}\ }\textbf {\bibinfo {volume} {35}},\ \bibinfo {pages} {311}
  (\bibinfo {year} {2003})}\BibitemShut {NoStop}%
\bibitem [{\citenamefont {Sun}\ \emph {et~al.}(2003)\citenamefont {Sun},
  \citenamefont {Salamon}, \citenamefont {Garnier},\ and\ \citenamefont
  {Averback}}]{Sun2003}%
  \BibitemOpen
  \bibfield  {author} {\bibinfo {author} {\bibfnamefont {Y.}~\bibnamefont
  {Sun}}, \bibinfo {author} {\bibfnamefont {M.~B.}\ \bibnamefont {Salamon}},
  \bibinfo {author} {\bibfnamefont {K.}~\bibnamefont {Garnier}}, \ and\
  \bibinfo {author} {\bibfnamefont {R.~S.}\ \bibnamefont {Averback}},\ }\href
  {\doibase 10.1103/PhysRevLett.91.167206} {\bibfield  {journal} {\bibinfo
  {journal} {Phys. Rev. Lett.}\ }\textbf {\bibinfo {volume} {91}},\ \bibinfo
  {pages} {167206} (\bibinfo {year} {2003})}\BibitemShut {NoStop}%
\bibitem [{\citenamefont {Lefloch}\ \emph {et~al.}(1992)\citenamefont
  {Lefloch}, \citenamefont {Hammann}, \citenamefont {Ocio},\ and\ \citenamefont
  {Vincent}}]{Lefloch1992}%
  \BibitemOpen
  \bibfield  {author} {\bibinfo {author} {\bibfnamefont {F.}~\bibnamefont
  {Lefloch}}, \bibinfo {author} {\bibfnamefont {J.}~\bibnamefont {Hammann}},
  \bibinfo {author} {\bibfnamefont {M.}~\bibnamefont {Ocio}}, \ and\ \bibinfo
  {author} {\bibfnamefont {E.}~\bibnamefont {Vincent}},\ }\href {\doibase
  10.1209/0295-5075/18/7/013} {\bibfield  {journal} {\bibinfo  {journal}
  {Europhys. Lett.}\ }\textbf {\bibinfo {volume} {18}},\ \bibinfo {pages} {647}
  (\bibinfo {year} {1992})}\BibitemShut {NoStop}%
\bibitem [{\citenamefont {Rodriguez-Carvajal}(1993)}]{Rodriguez-Carvajal1993}%
  \BibitemOpen
  \bibfield  {author} {\bibinfo {author} {\bibfnamefont {J.}~\bibnamefont
  {Rodriguez-Carvajal}},\ }\href {\doibase
  https://doi.org/10.1016/0921-4526(93)90108-I} {\bibfield  {journal} {\bibinfo
   {journal} {Physica B: Condensed Matter}\ }\textbf {\bibinfo {volume}
  {192}},\ \bibinfo {pages} {55 } (\bibinfo {year} {1993})}\BibitemShut
  {NoStop}%
\bibitem [{\citenamefont {Malinowski}\ \emph {et~al.}(2011)\citenamefont
  {Malinowski}, \citenamefont {Bezusyy}, \citenamefont {Minikayev},
  \citenamefont {Dziawa}, \citenamefont {Syryanyy},\ and\ \citenamefont
  {Sawicki}}]{Malinowski2011}%
  \BibitemOpen
  \bibfield  {author} {\bibinfo {author} {\bibfnamefont {A.}~\bibnamefont
  {Malinowski}}, \bibinfo {author} {\bibfnamefont {V.~L.}\ \bibnamefont
  {Bezusyy}}, \bibinfo {author} {\bibfnamefont {R.}~\bibnamefont {Minikayev}},
  \bibinfo {author} {\bibfnamefont {P.}~\bibnamefont {Dziawa}}, \bibinfo
  {author} {\bibfnamefont {Y.}~\bibnamefont {Syryanyy}}, \ and\ \bibinfo
  {author} {\bibfnamefont {M.}~\bibnamefont {Sawicki}},\ }\href {\doibase
  10.1103/PhysRevB.84.024409} {\bibfield  {journal} {\bibinfo  {journal} {Phys.
  Rev. B}\ }\textbf {\bibinfo {volume} {84}},\ \bibinfo {pages} {024409}
  (\bibinfo {year} {2011})}\BibitemShut {NoStop}%
\bibitem [{\citenamefont {Mulder}\ \emph {et~al.}(1982)\citenamefont {Mulder},
  \citenamefont {van Duyneveldt},\ and\ \citenamefont {Mydosh}}]{Mulder1982}%
  \BibitemOpen
  \bibfield  {author} {\bibinfo {author} {\bibfnamefont {C.~A.~M.}\
  \bibnamefont {Mulder}}, \bibinfo {author} {\bibfnamefont {A.~J.}\
  \bibnamefont {van Duyneveldt}}, \ and\ \bibinfo {author} {\bibfnamefont
  {J.~A.}\ \bibnamefont {Mydosh}},\ }\href {\doibase 10.1103/PhysRevB.25.515}
  {\bibfield  {journal} {\bibinfo  {journal} {Phys. Rev. B}\ }\textbf {\bibinfo
  {volume} {25}},\ \bibinfo {pages} {515} (\bibinfo {year} {1982})}\BibitemShut
  {NoStop}%
\bibitem [{\citenamefont {S\"ullow}\ \emph {et~al.}(1997)\citenamefont
  {S\"ullow}, \citenamefont {Nieuwenhuys}, \citenamefont {Menovsky},
  \citenamefont {Mydosh}, \citenamefont {Mentink}, \citenamefont {Mason},\ and\
  \citenamefont {Buyers}}]{Suellow1997}%
  \BibitemOpen
  \bibfield  {author} {\bibinfo {author} {\bibfnamefont {S.}~\bibnamefont
  {S\"ullow}}, \bibinfo {author} {\bibfnamefont {G.~J.}\ \bibnamefont
  {Nieuwenhuys}}, \bibinfo {author} {\bibfnamefont {A.~A.}\ \bibnamefont
  {Menovsky}}, \bibinfo {author} {\bibfnamefont {J.~A.}\ \bibnamefont
  {Mydosh}}, \bibinfo {author} {\bibfnamefont {S.~A.~M.}\ \bibnamefont
  {Mentink}}, \bibinfo {author} {\bibfnamefont {T.~E.}\ \bibnamefont {Mason}},
  \ and\ \bibinfo {author} {\bibfnamefont {W.~J.~L.}\ \bibnamefont {Buyers}},\
  }\href {\doibase 10.1103/PhysRevLett.78.354} {\bibfield  {journal} {\bibinfo
  {journal} {Phys. Rev. Lett.}\ }\textbf {\bibinfo {volume} {78}},\ \bibinfo
  {pages} {354} (\bibinfo {year} {1997})}\BibitemShut {NoStop}%
\bibitem [{\citenamefont {Mahendiran}\ \emph {et~al.}(2003)\citenamefont
  {Mahendiran}, \citenamefont {Br\'eard}, \citenamefont {Hervieu},
  \citenamefont {Raveau},\ and\ \citenamefont {Schiffer}}]{Mahendiran2003}%
  \BibitemOpen
  \bibfield  {author} {\bibinfo {author} {\bibfnamefont {R.}~\bibnamefont
  {Mahendiran}}, \bibinfo {author} {\bibfnamefont {Y.}~\bibnamefont
  {Br\'eard}}, \bibinfo {author} {\bibfnamefont {M.}~\bibnamefont {Hervieu}},
  \bibinfo {author} {\bibfnamefont {B.}~\bibnamefont {Raveau}}, \ and\ \bibinfo
  {author} {\bibfnamefont {P.}~\bibnamefont {Schiffer}},\ }\href {\doibase
  10.1103/PhysRevB.68.104402} {\bibfield  {journal} {\bibinfo  {journal} {Phys.
  Rev. B}\ }\textbf {\bibinfo {volume} {68}},\ \bibinfo {pages} {104402}
  (\bibinfo {year} {2003})}\BibitemShut {NoStop}%
\bibitem [{\citenamefont {Mydosh}(2015)}]{Mydosh2015}%
  \BibitemOpen
  \bibfield  {author} {\bibinfo {author} {\bibfnamefont {J.~A.}\ \bibnamefont
  {Mydosh}},\ }\href {\doibase 10.1088/0034-4885/78/5/052501} {\bibfield
  {journal} {\bibinfo  {journal} {Reports on Progress in Physics}\ }\textbf
  {\bibinfo {volume} {78}},\ \bibinfo {pages} {052501} (\bibinfo {year}
  {2015})}\BibitemShut {NoStop}%
\bibitem [{\citenamefont {Mydosh}(1993)}]{Mydosh1993}%
  \BibitemOpen
  \bibfield  {author} {\bibinfo {author} {\bibfnamefont {J.~A.}\ \bibnamefont
  {Mydosh}},\ }\href@noop {} {\emph {\bibinfo {title} {Spin Glasses: An
  Experimental Introduction}}}\ (\bibinfo  {publisher} {Taylor \& Francis},\
  \bibinfo {address} {London},\ \bibinfo {year} {1993})\BibitemShut {NoStop}%
\bibitem [{\citenamefont {Chatterjee}\ \emph {et~al.}(2009)\citenamefont
  {Chatterjee}, \citenamefont {Giri}, \citenamefont {De},\ and\ \citenamefont
  {Majumdar}}]{Chatterjee2009}%
  \BibitemOpen
  \bibfield  {author} {\bibinfo {author} {\bibfnamefont {S.}~\bibnamefont
  {Chatterjee}}, \bibinfo {author} {\bibfnamefont {S.}~\bibnamefont {Giri}},
  \bibinfo {author} {\bibfnamefont {S.~K.}\ \bibnamefont {De}}, \ and\ \bibinfo
  {author} {\bibfnamefont {S.}~\bibnamefont {Majumdar}},\ }\href {\doibase
  10.1103/PhysRevB.79.092410} {\bibfield  {journal} {\bibinfo  {journal} {Phys.
  Rev. B}\ }\textbf {\bibinfo {volume} {79}},\ \bibinfo {pages} {092410}
  (\bibinfo {year} {2009})}\BibitemShut {NoStop}%
\bibitem [{\citenamefont {Luo}\ \emph {et~al.}(2008)\citenamefont {Luo},
  \citenamefont {Zhao}, \citenamefont {Pan},\ and\ \citenamefont
  {Wang}}]{Luo2008}%
  \BibitemOpen
  \bibfield  {author} {\bibinfo {author} {\bibfnamefont {Q.}~\bibnamefont
  {Luo}}, \bibinfo {author} {\bibfnamefont {D.~Q.}\ \bibnamefont {Zhao}},
  \bibinfo {author} {\bibfnamefont {M.~X.}\ \bibnamefont {Pan}}, \ and\
  \bibinfo {author} {\bibfnamefont {W.~H.}\ \bibnamefont {Wang}},\ }\href
  {\doibase 10.1063/1.2827198} {\bibfield  {journal} {\bibinfo  {journal}
  {Appl. Phys. Lett.}\ }\textbf {\bibinfo {volume} {92}},\ \bibinfo {pages}
  {011923} (\bibinfo {year} {2008})}\BibitemShut {NoStop}%
\bibitem [{\citenamefont {Viswanathan}\ and\ \citenamefont
  {Kumar}(2009)}]{Viswanathan2009}%
  \BibitemOpen
  \bibfield  {author} {\bibinfo {author} {\bibfnamefont {M.}~\bibnamefont
  {Viswanathan}}\ and\ \bibinfo {author} {\bibfnamefont {P.~S.~A.}\
  \bibnamefont {Kumar}},\ }\href {\doibase 10.1103/PhysRevB.80.012410}
  {\bibfield  {journal} {\bibinfo  {journal} {Phys. Rev. B}\ }\textbf {\bibinfo
  {volume} {80}},\ \bibinfo {pages} {012410} (\bibinfo {year}
  {2009})}\BibitemShut {NoStop}%
\bibitem [{\citenamefont {Fisher}\ and\ \citenamefont
  {Huse}(1986)}]{Fisher1986}%
  \BibitemOpen
  \bibfield  {author} {\bibinfo {author} {\bibfnamefont {D.~S.}\ \bibnamefont
  {Fisher}}\ and\ \bibinfo {author} {\bibfnamefont {D.~A.}\ \bibnamefont
  {Huse}},\ }\href {\doibase 10.1103/PhysRevLett.56.1601} {\bibfield  {journal}
  {\bibinfo  {journal} {Phys. Rev. Lett.}\ }\textbf {\bibinfo {volume} {56}},\
  \bibinfo {pages} {1601} (\bibinfo {year} {1986})}\BibitemShut {NoStop}%
\bibitem [{\citenamefont {Dho}\ \emph {et~al.}(2002)\citenamefont {Dho},
  \citenamefont {Kim},\ and\ \citenamefont {Hur}}]{Dho2002}%
  \BibitemOpen
  \bibfield  {author} {\bibinfo {author} {\bibfnamefont {J.}~\bibnamefont
  {Dho}}, \bibinfo {author} {\bibfnamefont {W.~S.}\ \bibnamefont {Kim}}, \ and\
  \bibinfo {author} {\bibfnamefont {N.~H.}\ \bibnamefont {Hur}},\ }\href
  {\doibase 10.1103/PhysRevLett.89.027202} {\bibfield  {journal} {\bibinfo
  {journal} {Phys. Rev. Lett.}\ }\textbf {\bibinfo {volume} {89}},\ \bibinfo
  {pages} {027202} (\bibinfo {year} {2002})}\BibitemShut {NoStop}%
\bibitem [{\citenamefont {Hanasaki}\ \emph {et~al.}(2007)\citenamefont
  {Hanasaki}, \citenamefont {Watanabe}, \citenamefont {Ohtsuka}, \citenamefont
  {K\'ezsm\'arki}, \citenamefont {Iguchi}, \citenamefont {Miyasaka},\ and\
  \citenamefont {Tokura}}]{Hanasaki2007}%
  \BibitemOpen
  \bibfield  {author} {\bibinfo {author} {\bibfnamefont {N.}~\bibnamefont
  {Hanasaki}}, \bibinfo {author} {\bibfnamefont {K.}~\bibnamefont {Watanabe}},
  \bibinfo {author} {\bibfnamefont {T.}~\bibnamefont {Ohtsuka}}, \bibinfo
  {author} {\bibfnamefont {I.}~\bibnamefont {K\'ezsm\'arki}}, \bibinfo {author}
  {\bibfnamefont {S.}~\bibnamefont {Iguchi}}, \bibinfo {author} {\bibfnamefont
  {S.}~\bibnamefont {Miyasaka}}, \ and\ \bibinfo {author} {\bibfnamefont
  {Y.}~\bibnamefont {Tokura}},\ }\href {\doibase 10.1103/PhysRevLett.99.086401}
  {\bibfield  {journal} {\bibinfo  {journal} {Phys. Rev. Lett.}\ }\textbf
  {\bibinfo {volume} {99}},\ \bibinfo {pages} {086401} (\bibinfo {year}
  {2007})}\BibitemShut {NoStop}%
\bibitem [{\citenamefont {Nam}\ \emph {et~al.}(2000)\citenamefont {Nam},
  \citenamefont {Mathieu}, \citenamefont {Nordblad}, \citenamefont {Khiem},\
  and\ \citenamefont {Phuc}}]{Nam2000}%
  \BibitemOpen
  \bibfield  {author} {\bibinfo {author} {\bibfnamefont {D.~N.~H.}\
  \bibnamefont {Nam}}, \bibinfo {author} {\bibfnamefont {R.}~\bibnamefont
  {Mathieu}}, \bibinfo {author} {\bibfnamefont {P.}~\bibnamefont {Nordblad}},
  \bibinfo {author} {\bibfnamefont {N.~V.}\ \bibnamefont {Khiem}}, \ and\
  \bibinfo {author} {\bibfnamefont {N.~X.}\ \bibnamefont {Phuc}},\ }\href
  {\doibase 10.1103/PhysRevB.62.8989} {\bibfield  {journal} {\bibinfo
  {journal} {Phys. Rev. B}\ }\textbf {\bibinfo {volume} {62}},\ \bibinfo
  {pages} {8989} (\bibinfo {year} {2000})}\BibitemShut {NoStop}%
\bibitem [{\citenamefont {Bhattacharyya}\ \emph {et~al.}(2011)\citenamefont
  {Bhattacharyya}, \citenamefont {Giri},\ and\ \citenamefont
  {Majumdar}}]{Bhattacharyya2011}%
  \BibitemOpen
  \bibfield  {author} {\bibinfo {author} {\bibfnamefont {A.}~\bibnamefont
  {Bhattacharyya}}, \bibinfo {author} {\bibfnamefont {S.}~\bibnamefont {Giri}},
  \ and\ \bibinfo {author} {\bibfnamefont {S.}~\bibnamefont {Majumdar}},\
  }\href {\doibase 10.1103/PhysRevB.83.134427} {\bibfield  {journal} {\bibinfo
  {journal} {Phys. Rev. B}\ }\textbf {\bibinfo {volume} {83}},\ \bibinfo
  {pages} {134427} (\bibinfo {year} {2011})}\BibitemShut {NoStop}%
\bibitem [{\citenamefont {Maji}\ \emph {et~al.}(2011)\citenamefont {Maji},
  \citenamefont {Suresh},\ and\ \citenamefont {Nigam}}]{Maji2011}%
  \BibitemOpen
  \bibfield  {author} {\bibinfo {author} {\bibfnamefont {B.}~\bibnamefont
  {Maji}}, \bibinfo {author} {\bibfnamefont {K.~G.}\ \bibnamefont {Suresh}}, \
  and\ \bibinfo {author} {\bibfnamefont {A.~K.}\ \bibnamefont {Nigam}},\ }\href
  {http://stacks.iop.org/0953-8984/23/i=50/a=506002} {\bibfield  {journal}
  {\bibinfo  {journal} {Journal of Physics: Condensed Matter}\ }\textbf
  {\bibinfo {volume} {23}},\ \bibinfo {pages} {506002} (\bibinfo {year}
  {2011})}\BibitemShut {NoStop}%
\bibitem [{\citenamefont {Sasaki}\ \emph {et~al.}(2005)\citenamefont {Sasaki},
  \citenamefont {J\"onsson}, \citenamefont {Takayama},\ and\ \citenamefont
  {Mamiya}}]{Sasaki2005}%
  \BibitemOpen
  \bibfield  {author} {\bibinfo {author} {\bibfnamefont {M.}~\bibnamefont
  {Sasaki}}, \bibinfo {author} {\bibfnamefont {P.~E.}\ \bibnamefont
  {J\"onsson}}, \bibinfo {author} {\bibfnamefont {H.}~\bibnamefont {Takayama}},
  \ and\ \bibinfo {author} {\bibfnamefont {H.}~\bibnamefont {Mamiya}},\ }\href
  {\doibase 10.1103/PhysRevB.71.104405} {\bibfield  {journal} {\bibinfo
  {journal} {Phys. Rev. B}\ }\textbf {\bibinfo {volume} {71}},\ \bibinfo
  {pages} {104405} (\bibinfo {year} {2005})}\BibitemShut {NoStop}%
\bibitem [{\citenamefont {Fisher}\ and\ \citenamefont
  {Huse}(1988{\natexlab{a}})}]{Fisher1988}%
  \BibitemOpen
  \bibfield  {author} {\bibinfo {author} {\bibfnamefont {D.~S.}\ \bibnamefont
  {Fisher}}\ and\ \bibinfo {author} {\bibfnamefont {D.~A.}\ \bibnamefont
  {Huse}},\ }\href {\doibase 10.1103/PhysRevB.38.373} {\bibfield  {journal}
  {\bibinfo  {journal} {Phys. Rev. B}\ }\textbf {\bibinfo {volume} {38}},\
  \bibinfo {pages} {373} (\bibinfo {year} {1988}{\natexlab{a}})}\BibitemShut
  {NoStop}%
\bibitem [{\citenamefont {Fisher}\ and\ \citenamefont
  {Huse}(1988{\natexlab{b}})}]{Fisher1988a}%
  \BibitemOpen
  \bibfield  {author} {\bibinfo {author} {\bibfnamefont {D.~S.}\ \bibnamefont
  {Fisher}}\ and\ \bibinfo {author} {\bibfnamefont {D.~A.}\ \bibnamefont
  {Huse}},\ }\href {\doibase 10.1103/PhysRevB.38.386} {\bibfield  {journal}
  {\bibinfo  {journal} {Phys. Rev. B}\ }\textbf {\bibinfo {volume} {38}},\
  \bibinfo {pages} {386} (\bibinfo {year} {1988}{\natexlab{b}})}\BibitemShut
  {NoStop}%
\bibitem [{\citenamefont {Kundu}\ \emph {et~al.}(2019)\citenamefont {Kundu},
  \citenamefont {Dey}, \citenamefont {Prinz-Zwick}, \citenamefont {B\"uttgen},\
  and\ \citenamefont {Mahajan}}]{Kundu2019}%
  \BibitemOpen
  \bibfield  {author} {\bibinfo {author} {\bibfnamefont {S.}~\bibnamefont
  {Kundu}}, \bibinfo {author} {\bibfnamefont {T.}~\bibnamefont {Dey}}, \bibinfo
  {author} {\bibfnamefont {M.}~\bibnamefont {Prinz-Zwick}}, \bibinfo {author}
  {\bibfnamefont {N.}~\bibnamefont {B\"uttgen}}, \ and\ \bibinfo {author}
  {\bibfnamefont {A.~V.}\ \bibnamefont {Mahajan}},\ }\href {\doibase
  https://doi.org/10.1016/j.jmmm.2019.02.012} {\bibfield  {journal} {\bibinfo
  {journal} {Journal of Magnetism and Magnetic Materials}\ }\textbf {\bibinfo
  {volume} {481}},\ \bibinfo {pages} {77 } (\bibinfo {year}
  {2019})}\BibitemShut {NoStop}%
\bibitem [{\citenamefont {Gopal}(1966)}]{Gopal1966}%
  \BibitemOpen
  \bibfield  {author} {\bibinfo {author} {\bibfnamefont {E.~S.~R.}\
  \bibnamefont {Gopal}},\ }\href@noop {} {\emph {\bibinfo {title} {Specific
  Heats at Low Temperatures}}}\ (\bibinfo  {publisher} {Plenum Press},\
  \bibinfo {address} {New York},\ \bibinfo {year} {1966})\BibitemShut {NoStop}%
\bibitem [{\citenamefont {Kumar}\ \emph {et~al.}(2016)\citenamefont {Kumar},
  \citenamefont {Sheptyakov}, \citenamefont {Khuntia}, \citenamefont {Rolfs},
  \citenamefont {Freeman}, \citenamefont {R\o{}nnow}, \citenamefont {Dey},
  \citenamefont {Baenitz},\ and\ \citenamefont {Mahajan}}]{Kumar2016}%
  \BibitemOpen
  \bibfield  {author} {\bibinfo {author} {\bibfnamefont {R.}~\bibnamefont
  {Kumar}}, \bibinfo {author} {\bibfnamefont {D.}~\bibnamefont {Sheptyakov}},
  \bibinfo {author} {\bibfnamefont {P.}~\bibnamefont {Khuntia}}, \bibinfo
  {author} {\bibfnamefont {K.}~\bibnamefont {Rolfs}}, \bibinfo {author}
  {\bibfnamefont {P.~G.}\ \bibnamefont {Freeman}}, \bibinfo {author}
  {\bibfnamefont {H.~M.}\ \bibnamefont {R\o{}nnow}}, \bibinfo {author}
  {\bibfnamefont {T.}~\bibnamefont {Dey}}, \bibinfo {author} {\bibfnamefont
  {M.}~\bibnamefont {Baenitz}}, \ and\ \bibinfo {author} {\bibfnamefont
  {A.~V.}\ \bibnamefont {Mahajan}},\ }\href {\doibase
  10.1103/PhysRevB.94.174410} {\bibfield  {journal} {\bibinfo  {journal} {Phys.
  Rev. B}\ }\textbf {\bibinfo {volume} {94}},\ \bibinfo {pages} {174410}
  (\bibinfo {year} {2016})}\BibitemShut {NoStop}%
\bibitem [{\citenamefont {{Brando, M.}}\ \emph {et~al.}(2002)\citenamefont
  {{Brando, M.}}, \citenamefont {{B\"uttgen, N.}}, \citenamefont {{Fritsch,
  V.}}, \citenamefont {{Hemberger, J.}}, \citenamefont {{Kaps, H.}},
  \citenamefont {{Krug von Nidda, H.-A.}}, \citenamefont {{Nicklas, M.}},
  \citenamefont {{Pucher, K.}}, \citenamefont {{Trinkl, W.}}, \citenamefont
  {{Loidl, A.}}, \citenamefont {{Scheidt, E. W.}}, \citenamefont {{Klemm,
  M.}},\ and\ \citenamefont {{Horn, S.}}}]{Brando2002}%
  \BibitemOpen
  \bibfield  {author} {\bibinfo {author} {\bibnamefont {{Brando, M.}}},
  \bibinfo {author} {\bibnamefont {{B\"uttgen, N.}}}, \bibinfo {author}
  {\bibnamefont {{Fritsch, V.}}}, \bibinfo {author} {\bibnamefont {{Hemberger,
  J.}}}, \bibinfo {author} {\bibnamefont {{Kaps, H.}}}, \bibinfo {author}
  {\bibnamefont {{Krug von Nidda, H.-A.}}}, \bibinfo {author} {\bibnamefont
  {{Nicklas, M.}}}, \bibinfo {author} {\bibnamefont {{Pucher, K.}}}, \bibinfo
  {author} {\bibnamefont {{Trinkl, W.}}}, \bibinfo {author} {\bibnamefont
  {{Loidl, A.}}}, \bibinfo {author} {\bibnamefont {{Scheidt, E. W.}}}, \bibinfo
  {author} {\bibnamefont {{Klemm, M.}}}, \ and\ \bibinfo {author} {\bibnamefont
  {{Horn, S.}}},\ }\href {\doibase 10.1140/epjb/e20020033} {\bibfield
  {journal} {\bibinfo  {journal} {Eur. Phys. J. B}\ }\textbf {\bibinfo {volume}
  {25}},\ \bibinfo {pages} {289} (\bibinfo {year} {2002})}\BibitemShut
  {NoStop}%
\bibitem [{\citenamefont {Trinkl}\ \emph
  {et~al.}(2000{\natexlab{b}})\citenamefont {Trinkl}, \citenamefont
  {B\"uttgen}, \citenamefont {Kaps}, \citenamefont {Loidl}, \citenamefont
  {Klemm},\ and\ \citenamefont {Horn}}]{Trinkl2000}%
  \BibitemOpen
  \bibfield  {author} {\bibinfo {author} {\bibfnamefont {W.}~\bibnamefont
  {Trinkl}}, \bibinfo {author} {\bibfnamefont {N.}~\bibnamefont {B\"uttgen}},
  \bibinfo {author} {\bibfnamefont {H.}~\bibnamefont {Kaps}}, \bibinfo {author}
  {\bibfnamefont {A.}~\bibnamefont {Loidl}}, \bibinfo {author} {\bibfnamefont
  {M.}~\bibnamefont {Klemm}}, \ and\ \bibinfo {author} {\bibfnamefont
  {S.}~\bibnamefont {Horn}},\ }\href {\doibase 10.1103/PhysRevB.62.1793}
  {\bibfield  {journal} {\bibinfo  {journal} {Phys. Rev. B}\ }\textbf {\bibinfo
  {volume} {62}},\ \bibinfo {pages} {1793} (\bibinfo {year}
  {2000}{\natexlab{b}})}\BibitemShut {NoStop}%
\bibitem [{\citenamefont {Schmidt}\ \emph {et~al.}(2017)\citenamefont
  {Schmidt}, \citenamefont {Zimmer},\ and\ \citenamefont
  {Magalhaes}}]{Schmidt2017}%
  \BibitemOpen
  \bibfield  {author} {\bibinfo {author} {\bibfnamefont {M.}~\bibnamefont
  {Schmidt}}, \bibinfo {author} {\bibfnamefont {F.~M.}\ \bibnamefont {Zimmer}},
  \ and\ \bibinfo {author} {\bibfnamefont {S.~G.}\ \bibnamefont {Magalhaes}},\
  }\href {\doibase 10.1088/1361-648x/aa6060} {\bibfield  {journal} {\bibinfo
  {journal} {Journal of Physics: Condensed Matter}\ }\textbf {\bibinfo {volume}
  {29}},\ \bibinfo {pages} {165801} (\bibinfo {year} {2017})}\BibitemShut
  {NoStop}%
\bibitem [{\citenamefont {Chen}\ \emph {et~al.}(2014)\citenamefont {Chen},
  \citenamefont {Kee},\ and\ \citenamefont {Kim}}]{Chen2014}%
  \BibitemOpen
  \bibfield  {author} {\bibinfo {author} {\bibfnamefont {G.}~\bibnamefont
  {Chen}}, \bibinfo {author} {\bibfnamefont {H.-Y.}\ \bibnamefont {Kee}}, \
  and\ \bibinfo {author} {\bibfnamefont {Y.~B.}\ \bibnamefont {Kim}},\ }\href
  {\doibase 10.1103/PhysRevLett.113.197202} {\bibfield  {journal} {\bibinfo
  {journal} {Phys. Rev. Lett.}\ }\textbf {\bibinfo {volume} {113}},\ \bibinfo
  {pages} {197202} (\bibinfo {year} {2014})}\BibitemShut {NoStop}%
\bibitem [{\citenamefont {Chen}\ \emph {et~al.}(2016)\citenamefont {Chen},
  \citenamefont {Kee},\ and\ \citenamefont {Kim}}]{Chen2016}%
  \BibitemOpen
  \bibfield  {author} {\bibinfo {author} {\bibfnamefont {G.}~\bibnamefont
  {Chen}}, \bibinfo {author} {\bibfnamefont {H.-Y.}\ \bibnamefont {Kee}}, \
  and\ \bibinfo {author} {\bibfnamefont {Y.~B.}\ \bibnamefont {Kim}},\ }\href
  {\doibase 10.1103/PhysRevB.93.245134} {\bibfield  {journal} {\bibinfo
  {journal} {Phys. Rev. B}\ }\textbf {\bibinfo {volume} {93}},\ \bibinfo
  {pages} {245134} (\bibinfo {year} {2016})}\BibitemShut {NoStop}%
\bibitem [{\citenamefont {Chen}\ and\ \citenamefont {Lee}(2018)}]{Chen2018}%
  \BibitemOpen
  \bibfield  {author} {\bibinfo {author} {\bibfnamefont {G.}~\bibnamefont
  {Chen}}\ and\ \bibinfo {author} {\bibfnamefont {P.~A.}\ \bibnamefont {Lee}},\
  }\href {\doibase 10.1103/PhysRevB.97.035124} {\bibfield  {journal} {\bibinfo
  {journal} {Phys. Rev. B}\ }\textbf {\bibinfo {volume} {97}},\ \bibinfo
  {pages} {035124} (\bibinfo {year} {2018})}\BibitemShut {NoStop}%
\bibitem [{\citenamefont {Andrade}\ \emph {et~al.}(2018)\citenamefont
  {Andrade}, \citenamefont {Hoyos}, \citenamefont {Rachel},\ and\ \citenamefont
  {Vojta}}]{Andrade2018}%
  \BibitemOpen
  \bibfield  {author} {\bibinfo {author} {\bibfnamefont {E.~C.}\ \bibnamefont
  {Andrade}}, \bibinfo {author} {\bibfnamefont {J.~A.}\ \bibnamefont {Hoyos}},
  \bibinfo {author} {\bibfnamefont {S.}~\bibnamefont {Rachel}}, \ and\ \bibinfo
  {author} {\bibfnamefont {M.}~\bibnamefont {Vojta}},\ }\href {\doibase
  10.1103/PhysRevLett.120.097204} {\bibfield  {journal} {\bibinfo  {journal}
  {Phys. Rev. Lett.}\ }\textbf {\bibinfo {volume} {120}},\ \bibinfo {pages}
  {097204} (\bibinfo {year} {2018})}\BibitemShut {NoStop}%
\end{thebibliography}%

\end{document}